\documentclass[cup6b]{cupbook}

\usepackage{graphicx} 
\usepackage{color}
\usepackage{amsmath}
\usepackage{amssymb}

\begin{document}

%\maketitle
\cleardoublepage
% Dummy out the chapter number in the preprint, in part because it does not get centered.
\def\thechapter{}

\author[E. Athanassoula]{E. Athanassoula\\
Aix Marseille Universit\'e, CNRS,\\
LAM (Laboratoire d'Astrophysique de Marseille)\\
UMR 7326, 13388, Marseille, France\\
lia@oamp.fr} 

\chapter[Bars and secular evolution in disk galaxies]{Bars and secular evolution in disk galaxies:\\ Theoretical input}

%
%%%%%%%%%%%%%%%%%%%%%%%%%%%%%%%%%%%%%%%%%%%%%%%%%%%%%%%%%%%%%%%%%%%%%%%%%%%%%%%%
%

\abstract{Bars play a major role in driving the evolution of disk galaxies and
in shaping their present properties. They cause angular momentum to be
redistributed within the galaxy, emitted mainly from (near-)resonant material
at the inner Lindblad resonance of the bar, and absorbed mainly by
(near-)resonant material in the spheroid (i.e., the halo and, whenever relevant,
the bulge) and in the outer disk. Spheroids delay and slow down the initial
growth of the bar they host, but, at the later stages of the evolution, they
strengthen the bar by absorbing angular momentum. Increased velocity dispersion
in the (near-)resonant regions delays bar formation and leads to less strong
bars.\looseness-1

When bars form they are vertically thin, but soon their inner parts puff up and
form what is commonly known as the boxy/peanut bulge. This gives a complex and
interesting shape to the bar which explains a number of observations and also
argues that the \textit{COBE}/DIRBE bar and the Long bar in our Galaxy are,
respectively, the thin and the thick part of a single bar.

The value of the bar pattern speed may be set by optimising the balance between
emitters and absorbers, so that a maximum amount of angular momentum is
redistributed. As they evolve, bars grow stronger and rotate slower. Bars also
redistribute matter within the galaxy, create a disky bulge (pseudo-bulge),
increase the disk scale-length and extent and drive substructures such as
spirals and rings. They also affect the shape of the inner part of the spheroid,
which can evolve from spherical to triaxial.}\newpage

\def\thechapter{4}

%
%%%%%%%%%%%%%%%%%%%%%%%%%%%%%%%%%%%%%%%%%%%%%%%%%%%%%%%%%%%%%%%%%%%%%%%%%%%%%%%%
%

\section{Introductory remarks}
\label{sec:remarks}

In the $\Lambda$CDM model, galaxies are formed in dark matter haloes, and, at
early times, merge frequently with their neighbours. As time evolves (and
redshift decreases), the rate of mergers decreases and the evolution of galaxies
changes from being merger-driven to a more internally driven one. This change is
progressive and the transition is very gradual. Generally, the internally driven
evolution is on a much longer timescale than the merger-driven one. It is now
usually termed {\it secular} (for slow), a term introduced by Kormendy (1979),
who made in that paper the first steps in linking this evolution with galaxy
morphology.

In the sixties, and partly through the seventies as well, theoretical work on
galaxy dynamics was mainly analytical. The working hypothesis usually was that
potentials are steady-state, or quasi-steady-state. Thus, given a potential or
type of potential, theoretical work would follow the motions of individual
particles, or would study collective effects aiming for self-consistent
solutions, by following, e.g., the Boltzmann equation (Binney \& Tremaine 2008).
In this way, the basis of orbital structure theory was set and a considerable
understanding of many dynamical effects was obtained. The advent of numerical
simulations, however, made it clear that galaxies evolve with time, so that a
quasi-steady-state approach can not give the complete picture.     

Secular evolution was the general subject of this series of lectures, which were
given in November 2011 in the XXIII Canary Islands Winter School of
Astrophysics. My specific subject was bar-driven secular evolution and was
presented from the theoretical viewpoint, although I included in many places
comparisons with observations. In this written version I concentrate on a few
specific topics, such as the angular momentum redistribution within the galaxy,
the role of resonances in this redistribution, and its results on bar evolution
and boxy/peanut bulges. I will discuss elsewhere the effects of gas and of halo
triaxiality and clumpiness. The main tool I used was $N$-body simulations, and,
albeit to a somewhat lesser extent, analytic work and orbital structure theory.
It is only by coupling several independent approaches that the answer to complex
questions, such as the ones we have tackled, can be obtained. 

Introductory material, useful for a better appreciation of some aspects of bar
evolution, can be found in Binney \& Tremaine (2008), while further related
material can be found in the reviews by Athanassoula (1984 -- on spiral
structure), Contopoulos \& Grosb{\o}l (1989 -- on orbits), Sellwood \& Wilkinson
(1993 -- on bars), Kormendy \& Kennicutt (2004 -- on secular evolution) and
Athanassoula (2008a -- on boxy/peanut and disky bulges),  as well as in other
chapters of this book.

%--------------------------------------------------------------------
\begin{figure*}
\begin{center}
\includegraphics[scale=0.63]{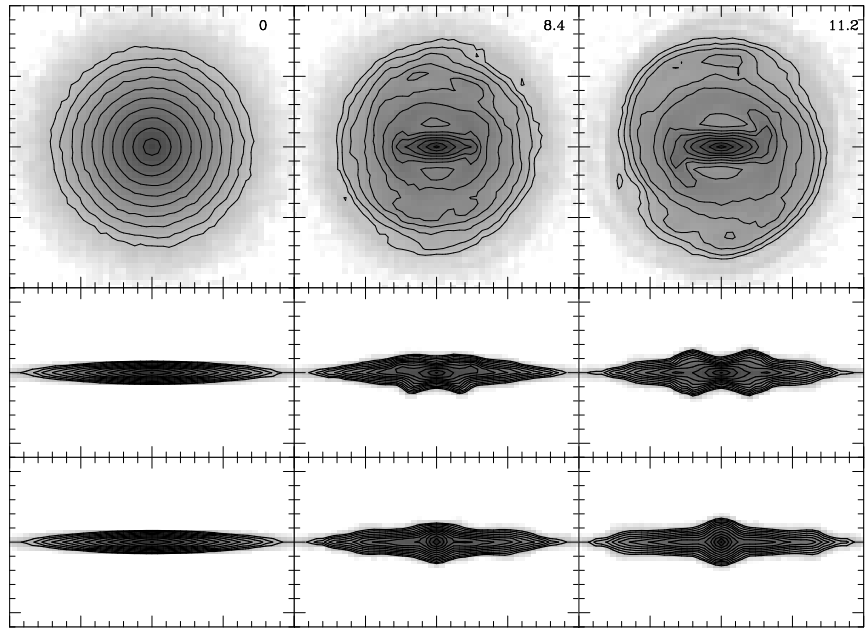}
\caption{Three snapshots showing the formation and evolution of a bar. Each
column corresponds to a given time (increasing from left to right), and each row
corresponds to a different viewing geometry, namely (from top to bottom)
face-on, side-on and end-on. The time, in Gyr, is given in the upper-right
corner of each upper panel. See text for further descriptions.} 
\label{fig:snapshot_evol}
\end{center}
\end{figure*}
%--------------------------------------------------------------------

%
%%%%%%%%%%%%%%%%%%%%%%%%%%%%%%%%%%%%%%%%%%%%%%%%%%%%%%%%%%%%%%%%%%%%%%%%%%%%%%%%
%

\section{Introduction}
\label{sec:intro}

$N$-body simulations have clearly shown that bars form spontaneously in galactic
disks. An example is given in Fig.~\ref{fig:snapshot_evol}, displaying the
face-on (upper panels), side-on\footnote{For the side-on view, the galaxy is
viewed edge-on, with the direction of the bar minor axis coinciding with the
line of sight.} (middle panels), and end-on\footnote{The end-on view  is also
edge-on, but now the line of sight coincides with the bar major axis.} (lower
panels) views of the disk component at three different times during the
formation and evolution. The left-hand panel shows the initial conditions of the
simulation, the right-hand one a snapshot at a time near the end of the
simulation, and the central panel a snapshot at an intermediate time. Before
plotting, I rotated the snapshots so that the major axis of the bar coincides
with the $x$ axis. 

Note that between the times of the central and right panels both the bar and the
disk have grown considerably in size, and that in both snapshots an inner ring
surrounds the bar. Note also that the initially thin disk becomes thick in the
inner parts. Seen side-on, it first becomes asymmetric with respect to the
equatorial plane and then puffs up to reach a peanut-like shape. Seen end-on, it
displays a bulge-like central concentration. From the face-on and the side-on
views we can infer that this concentration is simply the bar seen end-on. In a
real galaxy, however, where knowledge about the two other views would be
unavailable, this could be mistaken for a classical bulge, unless supplementary
photometric and/or kinematic information is available. Athanassoula (2005b) 
showed that this error could occur only if the angle between the bar major axis
and the line of sight was less that 5--10 degrees, i.e., within a rather
restricted range of viewing angles.  
 
Such bar formation and evolution processes had already been witnessed in the
pioneering $N$-body simulations of the early seventies and onward (e.g.,
Miller~\textit{et al.}~1970; Hohl 1971; Ostriker \& Peebles 1973; Sellwood 1980,
1981; Athanassoula \& Sellwood 1986; Sellwood \& Athanassoula 1986; Combes
\textit{et al.} 1990; Pfenniger \& Friedli 1991). Although technically these
simulations were not up to the level we are used to now (due to lower number of
particles, lower spatial and temporal resolution, absence or rigidity of the
halo component, a 2D geometry, etc.), they came to a number of interesting
results, two of which are closely related to what we will discuss here.

Ostriker \& Peebles (1973), using very simple simulations with only 150 to 500
particles, came to the conclusion that haloes can stabilise bars. This number of
particles is too low to describe adequately the bar-halo interaction and
particularly its effect on the bar growth. It is thus no surprise that their
result is partly flawed. Nevertheless, this paper, together with the subsequent
one by Ostriker \textit{et al.} (1974), gave a major impetus to research on dark
matter haloes, focusing both observational and theoretical effort on them.\looseness-1
 
Athanassoula \& Sellwood (1986), using 2D simulations with 40\,000 particles
only,  showed that bars grow slower in hotter disks (i.e., in disks with larger
velocity dispersions). They also confirmed a result which had been already found
in analytical mode calculations (e.g., Toomre 1981), namely that a higher
relative halo mass decreases the bar growth rate, so that bars grow {\it slower}
in disk galaxies with a larger $M_{\rm H}/M_{\rm D}$ ratio, where  $M_{\rm H}$
and $M_{\rm D}$ are the halo\footnote{There is some ambiguity in general about
what is meant by the term `halo mass'. In some cases it is the total halo
mass, but in others it is the mass within a radius encompassing the relevant
part of the simulated galaxy. In this case, since the simulations were 2D and
therefore the halo rigid, only a small-sized halo was considered, so the two
definitions coincide.} and disk masses, respectively. These results will be
discussed further in Section~\ref{subsec:bar-str}. 

%--------------------------------------------------------------------
\begin{figure*}
\begin{center}
\includegraphics[scale=0.68]{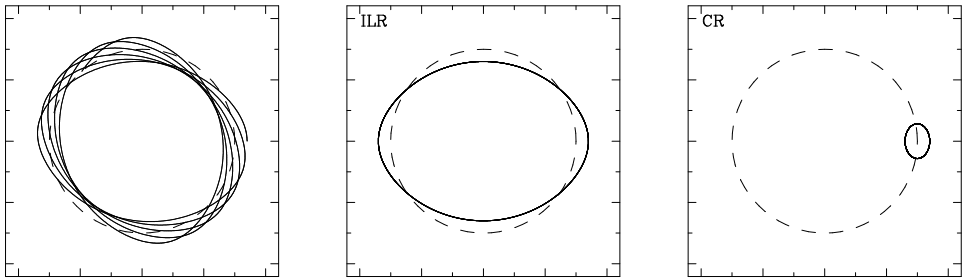}
\caption{Examples of epicyclic orbits. Left panel: A non-resonant orbit. Central
panel: Orbit at inner Lindblad resonance, i.e., for $(l,m)$\,=\,(-1,2). Right
panel: Orbit at corotation resonance, i.e., for $l$\,=\,0. In all three panels,
the dashed line gives the circular (guiding) orbit.}  
\label{fig:orbit_examples}
\end{center}
\end{figure*}
%--------------------------------------------------------------------

%
%%%%%%%%%%%%%%%%%%%%%%%%%%%%%%%%%%%%%%%%%%%%%%%%%%%%%%%%%%%%%%%%%%%%%%%%%%%%%%%%
%

\section{Orbits and resonances}
\label{sec:orbits}

Before starting on our quest for understanding the main bar formation and
evolution processes, let me first give a brief and considerably simplified
description of some basic notions of orbital structure theory. Readers
interested in more thorough and rigorous treatments can consult Arnold (1989)
and Lichtenberg \& Lieberman (1992).  

Let me consider a very simple potential composed of an axisymmetric part
(including all axisymmetric components) and a rigid bar rotating with a constant
angular velocity $\Omega_{\rm p}$. It is in general more convenient to work in a
frame of reference which co-rotates with the bar, in order to have a
time-independent potential (Binney \& Tremaine 2008) and I will simplify things
further by restricting myself to 2D motions. Any regular galactic orbit in this
potential\footnote{A number of concepts and results discussed in this section
are much more general, and can be applied to a more general class of
potentials.} can be characterised by two fundamental frequencies,
$\Omega_i,\,i$\,=\,1,2. In the epicyclic approximation these are $\Omega$, the
angular frequency of rotation around the galactic centre, and $\kappa$, the
epicyclic frequency, i.e., the frequency of radial oscillations. We say that  an
orbit is resonant if there are two integers $l$ and $m$ such that 

\begin{equation}
l\kappa+m(\Omega-\Omega_{\rm p})=0. 
\end{equation}

The most important resonances for our discussions here will be the Lindblad
resonances (inner and outer) and the corotation resonance. The inner Lindblad
resonance (hereafter ILR) occurs for $l$\,=\,$-1$ and $m$\,=\,2. Therefore, in a
frame of reference co-rotating with the bar, such orbits will close after one
revolution around the centre and two radial oscillations
(Fig.~\ref{fig:orbit_examples}).\linebreak Similarly, the outer Lindblad resonance
(hereafter OLR) occurs for $l$\,=\,1 and $m$\,=\,2. For $l$\,=\,0 we have the
corotation resonance (hereafter CR), where the angular frequency is equal to the
bar pattern speed, i.e., the particle co-rotates with the bar.   

Contrary to regular orbits, chaotic orbits (often also called irregular orbits)
do not have two fundamental frequencies and this property can be used to
distinguish them from regular orbits with the help of what is often called a
frequency analysis (Binney \& Spergel 1982; Laskar 1990). Let us also briefly
mention the so-called sticky orbits. Information on the dynamics and properties
of such orbits can be found in Contopoulos (2002). Here we will only mention
that, classified by eye, such orbits can be seen as being, say, regular over a
given interval of time and then, within a relatively short time, turning to
chaotic. Not too many years ago the existence and effect of non-regular orbits
on the structure and dynamics of galaxies was generally neglected, but it is
becoming progressively clear that this was wrong, so that such orbits are now
known to play a considerable role in many fields of galactic dynamics. 

\begin{figure*}
\begin{center}
\includegraphics[scale=0.74,angle=-90]{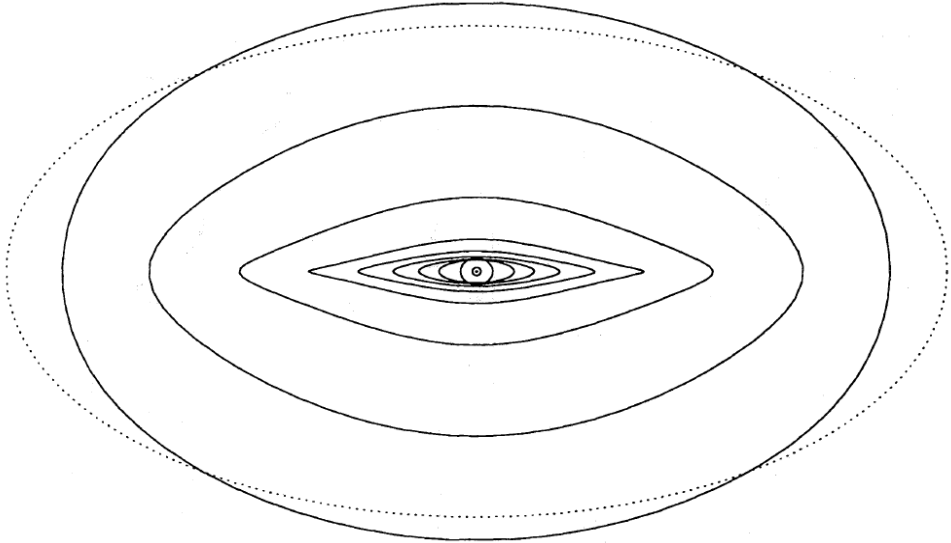}
\caption{Examples of orbits of the $x_1$ family. The outline of the bar is given
by a dashed line (reproduced from Athanassoula 1992a).} 
\label{fig:x1_examples}
\end{center}
\end{figure*}

By definition, resonant orbits close after a certain number of revolutions and a
certain number (not necessarily the same) of radial oscillations, and are often
referred to as periodic orbits. Several studies of such orbits in various bar
potentials have been made in 2D cases\footnote{3D cases will be discussed in
Section~\ref{subsec:BPorbits}.} (e.g., Contopoulos \& Papayannopoulos 1980;
Athanassoula \textit{et al.} 1983; Contopoulos \& Grosb{\o}l 1989). They show
that, in the equatorial plane, the main supporters of the bar are a family of
orbits elongated along the bar, named $x_1$ and having $l$\,=\,$-1$ and
$m$\,=\,2. Examples of members of this family can be seen in
Fig.~\ref{fig:x1_examples} here, or in Fig.~7 of Contopoulos \& Papayannopoulos
(1980), or Fig.~2 of Skokos \textit{et al.} (2002a). In most cases there is
another family of orbits with $l$\,=\,$-1$ and $m$\,=\,2, but which are oriented
perpendicularly to the bar and are named $x_2$. These play a crucial role in
determining the gas flow in the bar and the morphology of the inner kpc region
in the centre of the galaxy and will be discussed further by Isaac Shlosman
(this volume). Finally there are also two main families of periodic orbits at
CR, examples of which can be seen, e.g., in Fig.~3 and 4 of Contopoulos \&
Papayannopoulos (1980).

%--------------------------------------------------------------------
%--------------------------------------------------------------------

Periodic orbits can be stable or unstable and this can be tested by
consi\-dering another orbit very near the periodic one in phase space, i.e.,
with very similar values of positions and velocities. If the periodic orbit is
stable, then the new orbit will stay  in the immediate surroundings of the
periodic one and `wrap' itself around it. It can then be said that this new
orbit is `trapped' by the periodic one. Examples of trapped orbits can be seen
in Fig.~3.19 of Binney \& Tremaine (2008). The bar can then be considered as a
superposition of such orbits, trapped around members of the $x_1$ family, which
will thus be the backbone of the bar. On the other hand, if the periodic orbit
is unstable, then this second orbit will leave the vicinity of the periodic
orbit, and the distance between the two orbits in phase space will increase with
time, even though initially they were very near.   

The calculation of periodic orbits is straightforward, yet such orbits can
reveal crucial information on galactic structure and dynamics. A good example is
the work of Contopoulos (1980), who, with simple considerations on closed
orbits, was able to show that bars cannot extend beyond their CR. Further
work on periodic orbits coupled to hydrodynamic simulations gave an estimate of
the lower limit to the bar length, and the ratio $\mathcal{R}$ of the corotation
radius to the bar length was found to be in the range of 1.2\,$\pm$\,0.2
(Athanassoula 1992a, 1992b). Note, however, that the lower limit is only an
estimate, and not a strict limit as the upper limit. Nevertheless, several other
methods and works, including observational, gave results within the above-quoted
range, as reviewed by Elmegreen (1996) and by Corsini (2011). The bars for which
1.0\,$<$\,$\mathcal{R}$\,$<$\,1.4 are called {\it fast}, contrary to bars 
with $\mathcal{R}$\,$>$\,1.4, which are called {\it slow}. 
  
Finally, a straightforward superposition (with some smoothing) of stable
periodic orbits offers a very simple, yet most useful tool for studying
morphological or kinematical structures in disk galaxies and has been
successfully applied to bars, box/peanuts and rings (e.g., Patsis \textit{et
al.} 1997; Bureau \& Athanassoula 1999; Patsis \textit{et al.} 2002, 2003;
Patsis 2005; Patsis \textit{et al.} 2010).  

%
%%%%%%%%%%%%%%%%%%%%%%%%%%%%%%%%%%%%%%%%%%%%%%%%%%%%%%%%%%%%%%%%%%%%%%%%%%%%%%%%
%

\section{$N$-body simulations}
\label{sec:N-body}

The $N$-body simulations that we will discuss were tailored specifically for the
understanding of bar formation and evolution in a gas-less disk embedded in a
spherical spheroid. That is, the initial conditions were built so as to exclude,
in as much as possible, other instabilities, thus allowing us to focus on the
bar. Such initial conditions are often called dynamical (because they allow us
to concentrate on the dynamics), or simplified, controlled, or idealised
(because they exclude other effects so as to focus best on the one under study).
They allow us to make `sequences' of models, in which we vary only one parameter
and keep all the others fixed. For example, it is thus possible to obtain a
sequence of models with initially identical spheroids and identical disk density
profiles, but different velocity dispersions in the disk. 

The alternative to these simulations is cosmological simulations, and, more
specifically, zoom re-imulations. In such re-simulations a specific halo (or
galaxy), having the desired properties, is chosen from the final snapshot of a
full cosmological simulation. The simulation is then rerun with a higher
resolution for the parts which end up in the chosen galaxy or which come to a
close interaction with it, and also after having replaced a fraction of the dark
matter particles in those parts by gas particles. 

Zoom simulations are more general than the dynamical ones because the former
include all the effects that dynamical simulations have, deliberately,
neglected. However, they do not allow us to build sequences of models and also
have less resolution than the dynamical ones and necessitate much more computer
time and memory. Furthermore, some care is necessary because cosmological
simulations are known to have a few problems when compared with nearby galaxy
observations, concerning, e.g., the number and distribution of satellites, the
inner halo radial density profile, the formation of bulge-less galaxies, or the
Tully-Fisher relation (see, e.g., Silk \& Mamon 2012 for a review). Thus, the
zoom re-simulations could implicitly contain some non-realistic properties,
which are not in agreement with what is observed in nearby galaxies, and
therefore reach flawed results. Moreover, since many effects take place
simultaneously, it is often difficult to disentangle the contribution of each
one separately, which very strongly hampers the understanding of a phenomenon.
For example, it is impossible to fully understand the bar formation instability
if the model galaxy in which it occurs is continuously interacting or merging
with other galaxies. A more appropriate way would be to first understand the
formation and evolution of bars in an isolated galaxy, and then understand the
effect of the interactions and mergings as a function of the properties of the
intruder(s).

Thus, zoom simulations should not yet be considered as a replacement of
dynamical simulations, but rather as an alternative approach, allowing
comparisons with dynamical simulations after the basic instabilities has been
understood. A few studies using cosmological zoom simulations have been already
made and have given interesting results on the formation and properties of bars
(Romano-D\'iaz \textit{et al.} 2008; Scannapieco \& Athanassoula 2012; Kraljic
\textit{et al.} 2012). 

A non-trivial issue about dynamical $N$-body simulations is the creation of the
initial conditions. These assume that the spheroid and the disk are already in
place and, most important, that they are in equilibrium. This is very important,
since a system which is not in equilibrium will undergo violent relaxation and
transients, which can have undesirable secondary effects, such as spurious
heating of the disk or altering of its radial density profile. At least three
different classes of methods to create initial conditions have been developed so
far.\medskip

\begin{enumerate}[(a)]\listsize
\renewcommand{\theenumi}{(\alph{enumi})}

\item A wide variety of methods are based on Jeans's theorem (e.g., Zang 1976;
Athanassoula \& Sellwood 1986; Kuijken \& Dubinski 1995; Widrow \& Dubinski
2005; McMillan \& Dehnen 2007), or on Jeans's equations (e.g., Hernquist 1993).
In the case of multi-component systems, e.g., galaxies with a disk, a bulge and
a halo, the components are built separately and then either simply superposed
(e.g., Hernquist 1993), or the potential of the one is adiabatically grown in
the other before superposition (e.g., Barnes 1988; Shlosman \& Noguchi 1993;
Athanassoula 2003, 2007; McMillan \& Dehnen 2007). The former can be dangerous,
as the resulting model can be considerably off equilibrium. The latter is
strongly preferred to it, but still has the disadvantage that the adiabatic
growing of one component can alter the density profiles of the others, which is
not desirable when one wishes to make sequences of models. It also is not
trivial to device a method for assigning the velocities to the disk particles
without relying on the epicyclic approximation (but see Dehnen 1999). Last but
not least, this class of methods is not useful for complex systems such as
triaxial bulges or haloes.

\item The Schwarzschild method (Schwarzschild 1979) can also be used for
making initial conditions, but has been hardly used for this, because the
application is rather time consuming and not necessarily straightforward.

\item A very promising method for constructing equilibrium phase models for
stellar systems is the iterative method (Rodionov \textit{et al.} 2009). It
relies on constrained, or guided, evolution, so that the equilibrium solution
has a number of desired parameters and/or constraints. It is very powerful, to a
large extent due to its simplicity. It can be used for mass distributions with
an arbitrary geometry and a large variety of kinematical constraints. It has no
difficulty in creating triaxial spheroids, and the disks it creates do not
follow the epicyclic approximation, unless this has been imposed by the user. It
has lately been extended to include a gaseous component (Rodionov \&
Athanassoula 2011). Its only disadvantage is that it is computer intensive, so
that in some cases the time necessary to make the initial conditions is a
considerable fraction of the simulation time.

\end{enumerate}  

I would also like to stress here a terminology point which, although not limited
to simulations, is closely related to them. In general the dynamics of haloes
and bulges are very similar, with of course quantitative differences due to
their respective extent, mass and velocity dispersion values. For this reason, I
will use sometimes in these lecture notes the terms `halo' and `bulge'
specifically, while in others I will use the word `spheroid' in a generic way,
to designate the halo and/or the bulge component. The reasons for this are
sometimes historic (i.e., how it was mentioned in the original paper), or
quantitative (e.g., if the effect of the halo is quantitatively much stronger
that that of the bulge), or just for simplicity. The reader can mentally
interchange the terms as appropriate.

%
%%%%%%%%%%%%%%%%%%%%%%%%%%%%%%%%%%%%%%%%%%%%%%%%%%%%%%%%%%%%%%%%%%%%%%%%%%%%%%%%
%

\section{On angular momentum exchange and the role of resonances:\\ the analytic approach}
\label{sec:analytic}

Two papers are the pillars of the analytical work on angular momentum
redistribution in disk galaxies -- namely Lynden-Bell \& Kalnajs (1972) and
Tremaine \& Weinberg (1984) -- while further useful information can be found in,
e.g., Kalnajs (1971), Dekker (1974), Weinberg (1985, 1994), Athanassoula (2003),
Fuchs (2004), Fuchs \& Athanassoula (2005).  

In order to reach tractable analytic expressions, it is necessary to consider
the disk and the spheroid components separately, and use different
approximations in the two cases. For the disk we can use the epicyclic
approximation (i.e., we will assume that the disk orbits can be reasonably well
approximated by epicycles), while for the spheroid we will assume that the
distribution function depends only on the energy, as is the case for spherical
isotropic systems. The main results obtained in the papers listed above are:

\begin{enumerate}[(a)]\listsize
\renewcommand{\theenumi}{(\alph{enumi})}

\item Angular momentum is emitted or absorbed mainly at resonances. It is,
however, also possible to emit or absorb away from resonances if the potential
is not stationary, but grows or decays with time. Nevertheless, the contribution
of the non-resonant material to the total emission or absorption should remain
small, unless the growth or decay of the potential is important. 

\item In the disk component, angular momentum is emitted from the ILR and at
other $l~<~0$ resonances and absorbed at the OLR and at other $l~>~0$
resonances. It is also absorbed at CR, but, all else being equal, at lesser
quantities than at the Lindblad resonances.  
 
\item The spheroid absorbs angular momentum at all its resonances.

\item The global picture is thus that angular momentum is emitted from the bar
region and absorbed by the CR and OLR in the disk, and by all resonances in the
spheroid. Thus, angular momentum is transported from the inner parts of the
disk, to the part of the disk outside CR and to the spheroid resonant regions. 

\item For both the disk and the spheroid components it is possible to show
that, for the same perturbing potential and the same amount of resonant
material, a given resonance will emit or absorb more angular momentum if the
material there is colder (i.e., has a lower velocity dispersion). Therefore,
since the disk is always colder than the spheroid, it will absorb more angular
momentum per unit resonant mass. Nevertheless, the spheroid is much more massive
than the outer disk, so the amount of angular momentum it absorbs may exceed
that absorbed by the outer disk. 

\item Since the bar is inside corotation, it has negative energy and angular
momentum and as it emits angular momentum it gets destabilised, i.e., it grows
stronger. It is thus expected that the more angular momentum is emitted, the 
stronger the bar will become.

\end{enumerate}

%
%%%%%%%%%%%%%%%%%%%%%%%%%%%%%%%%%%%%%%%%%%%%%%%%%%%%%%%%%%%%%%%%%%%%%%%%%%%%%%%%
%
 
\section{On angular momentum exchange and the role of resonances:\\ input from simulations}
\label{sec:angmom}

\subsection{General comments}
\label{subsec:angmom-gen}

It is not possible to compare the analytical work mentioned in the previous
section directly with observations, because each galaxy is observed only at a
single time during its evolution, and neither angular momentum exchange nor
individual orbits can be directly observed. One should thus include an
intermediate step in the comparisons, namely $N$-body simulations. In these, it
is possible to follow directly not only the evolution in time, but also the
angular momentum exchange and the individual orbits, i.e., it is possible to
make direct comparisons of simulations with analytical work. Furthermore, one
can `observe' the simulation results using the same methods as for real galaxies
and make comparisons (Section~\ref{sec:observations}). Simulations thus provide
a meaningful and necessary link between analytical work and observations. 

In order to show that the analytical results discussed in
Section~\ref{sec:analytic} do apply to simulations it is necessary to go through 
a number of intermediate steps, i.e., to show  

\begin{enumerate}[(a)]\listsize
\renewcommand{\theenumi}{(\alph{enumi})}

\item
that there is a reasonable amount of mass at (near-)resonance both for the
disk and the spheroid components, 

\item
that angular momentum is emitted from the resonances in the bar region
and absorbed by all the spheroid resonances and the outer disk resonances, 

\item
that the contribution of the spheroid in the angular momentum
redistribution is important, 

\item
that, as a result of this angular momentum transfer, the bar
becomes stronger and slows down, 

\item
that stronger bars are found in simulations in which more angular
momentum has been exchanged within the galaxy, 

\item
and that more (less) angular momentum can be exchanged
when the emitting or absorbing material is colder (hotter).

\end{enumerate}

This sequence of steps was followed in two papers (Athanassoula 2002, hereafter
A02 and Athanassoula 2003, hereafter A03) whose techniques and results I will
review in the next subsections, giving, whenever useful, more extended
information (particularly on the techniques) than in the original paper, so that
the work can be easier followed by students and non-specialists. 

\subsection{Calculating the orbital frequencies}
\label{subsec:frequencies}

Our first step will be to calculate the fundamental orbital frequencies. Since
we are interested in the redistribution of $L_z$, we will focus on the angular
and the epicyclic frequency. The epicyclic frequency $\kappa$ can be calculated
with the help of the frequency analysis technique (Binney \& Spergel 1982;
Laskar 1990), which relies on a Fourier analysis of, e.g., the cylindrical
radius $R(t)$ along the orbit. The desired frequency is then obtained as the
frequency of the highest peak in the Fourier transform. The angular frequency 
$\Omega$ is more difficult to estimate, and in A02 and A03 I supplemented
frequency analysis with other methods, such as following the angle with time.  

Several technical details are important for the frequency analysis. It is
necessary to use windowing before doing the Fourier analysis, to improve the
accuracy. It is also necessary to keep in mind that some of the peaks of the
power spectrum are not independent frequencies, but simply harmonics of the
individual fundamental frequencies, or their combination. Furthermore, if one
needs considerable accuracy, one has to worry about the fact that in standard
Fast Fourier Transforms the step $d\omega$ between two adjacent frequencies is
constant, while the fundamental frequencies $\Omega_i$ will not necessarily fall
on a grid point. Except for the inaccuracy thus introduced, this will complicate
the handling of the harmonics. 

Frequency analysis can be applied to orbits in any analytic stationary galactic
potential, thus allowing the full calculation of the resonances and their
occupation (e.g., Papaphilipou \& Laskar 1996, 1998; Carpintero \& Aguilar 1998;
Valluri \textit{et al.} 2112). Contrary to such potentials, however, simulations
include full time evolution, so that the galactic potential, the bar pattern
speed, as well as the basic frequencies $\Omega$ and $\kappa$ of any orbit are
time-dependent. Thus, strictly speaking, the spectral analysis technique can not
be applied, at least as such. 

It is, nevertheless, possible to estimate the frequencies of a given orbit at
any given time $t$ by using the potential and bar pattern speed at this time $t$
(which are thus considered as frozen), as I did in A02 and A03. After freezing
the potential, I chose a number of particles at random from each component of
the simulation and calculated their orbits in the frozen potential, using as
initial conditions the positions and velocities of the particles in the
simulation at time $t$. It is necessary to take a sufficient number of particles
(of the order of 100\,000) in order to be able to define clearly the main
spectral lines. It is also necessary to follow the orbit for a sufficiently long
time (e.g., 40 orbital rotation patterns), in order to obtain narrow lines in the
spectrum. By doing so I do not assume that the potential stays unevolved over
such a long time. What I describe here just amounts to linking the properties of
a small part of the orbit calculated in the evolving simulation potential
(hereafter simulation orbit) to an equivalent part of the corresponding orbit
calculated in the frozen potential. The frequencies are then calculated for the
orbit in the frozen potential and attributed to the small part of the 
simulation orbit in question (and not to the whole of the simulation orbit).
This technique makes it possible to apply the frequency analysis method, as
described in A02, A03 and above, and thus to obtain the main frequencies of each
orbit at a given time. It is, furthermore, possible to follow the evolution by
choosing a number of snapshots during the simulation and performing the above
exercise separately for each one of them. The evolution can then be witnessed
from the sequence of the results, one for each chosen time. 

%--------------------------------------------------------------------
\begin{figure*}
\begin{center}
\includegraphics[scale=0.74]{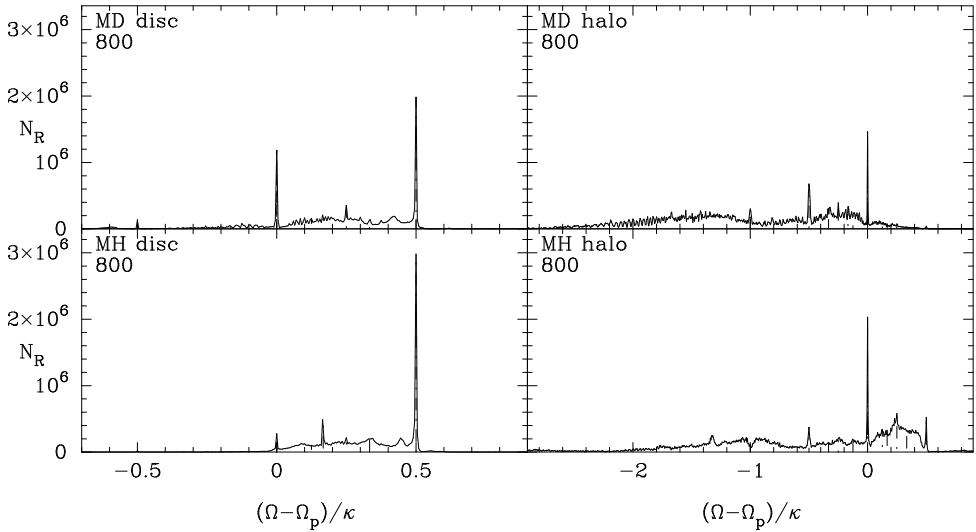}
\caption{Number density, $N_{\rm R}$, of particles as a function of the
frequency ratio $(\Omega-\Omega_{\rm p})/\kappa$, for two different simulations
(upper and lower panels, respectively) at a time near the end of the simulation.
The left panels correspond to the disk component and the right ones to the halo.
This figure is reproduced from A02.}
\label{fig:resonances}
\end{center}
\end{figure*}
%--------------------------------------------------------------------
 
\subsection{Material at resonance}
\label{subsec:spectrum}

Having calculated the fundamental frequencies as described in the previous
section, it is now possible to plot histograms of the number of particles -- or
of their total mass, if particles of unequal mass are used in the simulation --
as a function of the ratio of their frequencies measured in a frame of reference
co-rotating with the bar, i.e., as a function of $(\Omega-\Omega_{\rm
p})/\kappa$. This can be carried out separately for the particles describing the
various components, i.e., the disk, the halo, and the bulge. It was first
carried out in A02 and the results, for two different simulations, are shown in
Fig.~\ref{fig:resonances}.

Before making this histogram, it is necessary to eliminate chaotic orbits. Their
spectra differ strongly from those of regular orbits, consisting of a very large
number of non-isolated lines. They of course always have a `highest peak', but
this has no physical significance and is not a fundamental frequency of the
orbit. Eliminating chaotic orbits is non-trivial because of the existence of
sticky orbits (see Section~\ref{sec:orbits}) for which the results of the
classification as regular or chaotic may well depend on the chosen integration
time. Thus, although for regular orbits it is recommended to use a long
integration time in order to obtain narrow, well defined spectral peaks, for
sticky orbits integration times must be of the order of the characteristic
timescale of the problem. For instance, if the sticky orbit shifts from regular
to chaotic only after an integration time of the order of say ten Hubble times,
it will be of no concern to galactic dynamic problems and this orbit can for all
practical purposes be considered as regular.

It is clear from Fig.~\ref{fig:resonances} that the distribution of particles in
frequency is not homogeneous. In fact it has a few very strong peaks and a
number of smaller ones. The peaks are not randomly distributed; they are located
at the positions where the ratio $(\Omega-\Omega_{\rm p})/\kappa$ is equal to
the ratio of two integers, i.e., when the orbit is resonant and closes after a
given number of rotations and radial oscillations. The highest peak is at
$(\Omega-\Omega_{\rm p})/\kappa$\,=\,0.5, i.e., at the ILR. A second important
peak is located at $\Omega$\,=\,$\Omega_{\rm p}$, i.e., at CR where the particle
co-rotates with the bar. 

Other peaks, of lesser relative height, can be seen at other resonances, such as
the $-1/2$ (OLR), the 1/4 (often referred to as the ultraharmonic resonance --
UHR), the 1/3, the 2/3, etc. In all runs with a strong bar the ILR peak
dominates, as expected. But the height of these peaks differs from one
simulation to another and even from one time to another in the same simulation. 
  
This richness of structures in the resonance space could have been expected for
the disk component. What, however, initially came as a surprise was the
existence of strong resonant peaks in the spheroid. Two examples can be seen in
the right-hand panels of Fig.~\ref{fig:resonances}. In both, the strongest peak
is at corotation, and other peaks can be clearly seen at ILR, at
$(\Omega-\Omega_{\rm p})/\kappa$\,=\,$-0.5$ (OLR) and at other resonances. As
was the case for the disk, the absolute and relative heights of the peaks differ
from one simulation to another, as well as with time.

Thus, the results of A02 that we have discussed in this section show that, both
for the disk and the spheroid component, a very large fraction of the simulation
particles is at (near-)resonance. Note that this result is backed by a large
number of simulations. I have analysed the orbital structure and the resonances
of some 50 to 100 simulations and for a number of times per simulation. The
results of these, as yet unpublished, analyses are in good qualitative agreement
with what was presented and discussed in A02, A03 and here. 

Further confirmation was brought by a number of subsequent and independent
analyses (Mart\'inez-Valpuesta \textit{et al.} 2006; Ceverino \& Klypin 2007;
Dubinski \textit{et al.} 2009; Wozniak \& Michel-Dansac 2009; Saha \textit{et al.}
2012). These studies include many different models, with very different spheroid
mass profiles or distribution functions, as well as disks with different
velocity dispersions. Also different simulation codes were used, including the
Marseille GRAPE-3 and GRAPE-5 codes (Athanassoula \textit{et al.} 1998), GyrFalcon
(Dehnen 2000, 2002), FTM (Heller \& Shlosman 1994, Heller 1995), ART (Kravtsov
\textit{et al.} 1997), Dubinski's treecode (Dubinski 1996) and GADGET (Springel
\textit{et al.} 2001, Springel 2005).

% Note also that Ceverino \& Klypin (2007) have used a somewhat different
% approach, not freezing the potential before calculating the orbits. Instead they
% followed the particle orbits through a part of the simulation during which the
% galaxy potential (more specifically the bar potential and pattern speed) do not
% change too much. One can either be strict in defining the potential changes that
% can be tolerated in this criterion, in  which case the time over which one can
% integrate is very short, or lenient, but then the potential varies during the
% orbital calculation. Either way the peaks will be broad, but sufficiently
% well-defined to confirm the A02 results. Note also that this version of the
% frequency analysis is not suitable for deciding whether a given orbit is regular
% or chaotic, because the potential in any $N$-body simulation is by necessity
% somewhat noisy  and thus the number of chaotic orbits will be spuriously
% enhanced (Vasiliev \& Athanassoula 2012). It is thus not trivial to eliminate
% the non-regular orbits.  

Note also that Ceverino \& Klypin (2007) have used a somewhat different
approach, and did not freeze the potential before calculating the orbits.
Instead, they followed the particle orbits through a part of the simulation
during which the galaxy potential (more specifically the bar potential and
pattern speed) do not change too much. In this way they obtain a power spectrum
with much broader peaks than in the studies that analyse the orbits in a
sequence of frozen potentials. Nevertheless, the peaks are well-defined and
confirm the main A02 results -- namely that there are located at the main
resonances -- without the use of potential freezing. Note also that this version
of the frequency analysis is not suitable for deciding whether a given orbit is
regular or chaotic, but is considerably faster in computer time than the one
relying on a sequence of frozen potentials. 

\subsection{Angular momentum exchange}
\label{subsec:angmom}

In A03 I used $N$-body simulations to show that angular momentum is emitted at
the resonances within CR, i.e., in the bar region, and that it is absorbed at
resonances either in the spheroid, or in the disk from the CR outwards, as
predicted by analytic calculations. For this I calculated the angular momentum
of all particles in the simulation at two chosen times $t_1$ and $t_2$ and
plotted their difference, $\Delta J\,=\,J_{2}-J_{1}$ as a function of the
frequency ratio $(\Omega-\Omega_{\rm p})/\kappa$ of the particle orbit at time
$J_2$. An example of the result can be seen in Fig.~1 of A03. Note that
particles in the disk with a positive frequency ratio and particularly
particles at ILR have $\Delta J<0$, i.e., they emit angular momentum. On the
contrary particles in the spheroid have $\Delta J>0$, i.e. they absorb angular
momentum and particularly at the CR, followed by the ILR and OLR. Further
absorption can be seen at the disk CR, but it is considerably less than the
amount absorbed by the spheroid. The amount of angular momentum emitted or
absorbed at a given resonance is of course both model- and time-dependent, as
were the heights of the resonant peaks  (Section~\ref{subsec:spectrum}). On the
contrary, whether a given resonance absorbs or emits is model-independent, and
in good agreement with analytic predictions (Section~\ref{sec:analytic}). 

Thinking of the bar as an ensemble of orbits, it becomes clear that there are
many ways in which angular momentum can be lost from the bar region. The first
possibility is that the orbits in the bar, and therefore the bar itself, will
become more elongated. The second one is that orbits initially on circular
orbits closely outside the bar region will loose angular momentum and become
elongated and part of the bar, which will thus get longer and more massive. In
both cases the bar will become stronger in the process. The third alternative is
that the bar will rotate slower, i.e., its pattern speed will decrease. These
three possibilities were presented and discussed in A03, where it was shown that
that they are linked and occur concurrently. Thus evolution should make bars
longer, and/or more elongated, and/or more massive and/or slower rotating (A03).
Simulations agree fully with these predictions and go further, establishing that
all these occur concurrently, but not necessarily at the same pace. 

\subsection{Types of models}
\label{subsec:models}
 
\subsubsection{Models with maximum and models with sub-maximum disks}
\label{subsubsec:MH-MD}

Since the spheroid plays such a crucial role in the angular momentum
redistribution within the galaxy, it must also play a crucial role in the
formation and evolution of the bar. Athanassoula \& Misiriotis (2002, hereafter
AM02) tested this by analysing the bar properties in two very different types of
simulations, which they named MH (for Massive Halo) and MD (for Massive Disc),
respectively. Both types have a halo with a core, which is big in the MD types
and small in the MH ones. Thus, in MH models, the halo plays a substantial role
in the dynamics within the inner four or five disk scale lengths, while not
being too hot, so as not to impede the angular momentum absorption. On the
contrary, in MD models the disk dominates the dynamics within that radial
range. 

\begin{figure*}
\begin{center}
\includegraphics[scale=0.9, angle=-90]{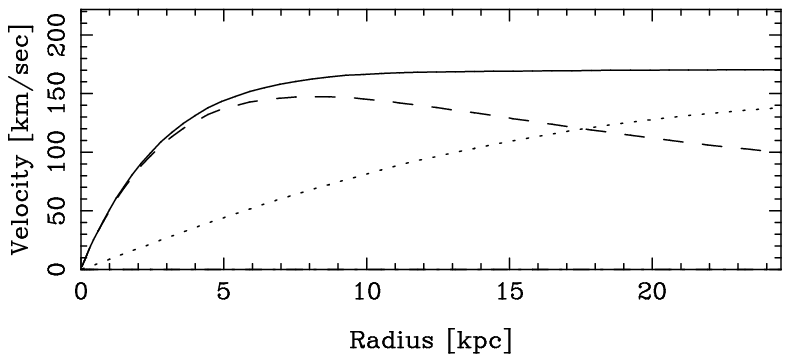}
\includegraphics[scale=0.9, angle=-90]{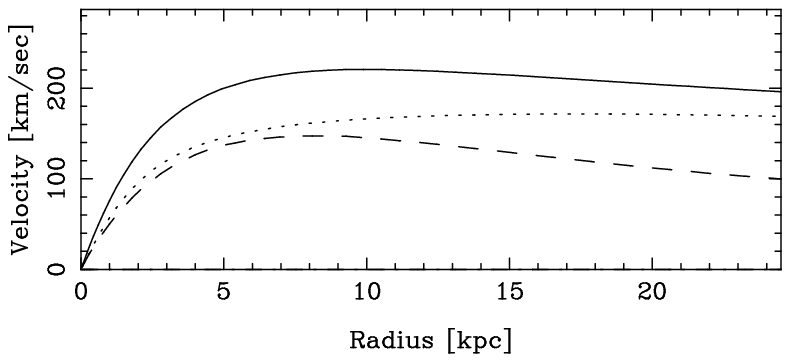}
\caption{Circular velocity curves of the initial condition of two models, one of
type MD (upper panel) and the other of type MH (lower panel). The solid line
gives the total circular velocity, while the dashed and the dotted ones give the
contributions of the disk and halo, respectively.}
\label{fig:rotcurves}
\end{center}
\end{figure*}

The circular velocity curves of these two types of models are compared in
Fig.~\ref{fig:rotcurves}. For the MD model (upper panel) the disk dominates the
dynamics in the inner few disk scale lengths, while this is not the case for the
MH model. MD-type models are what the observers call maximum disk models, while
the MH types have sub-maximum disks. It is not yet clear whether disks in real
galaxies are maximum, or sub-maximum, because different methods reach different
conclusions, as reviewed, e.g., by Bosma (2002). 

%--------------------------------------------------------------------
%--------------------------------------------------------------------

%--------------------------------------------------------------------
\begin{figure*}
\begin{center}
\includegraphics[scale=0.76]{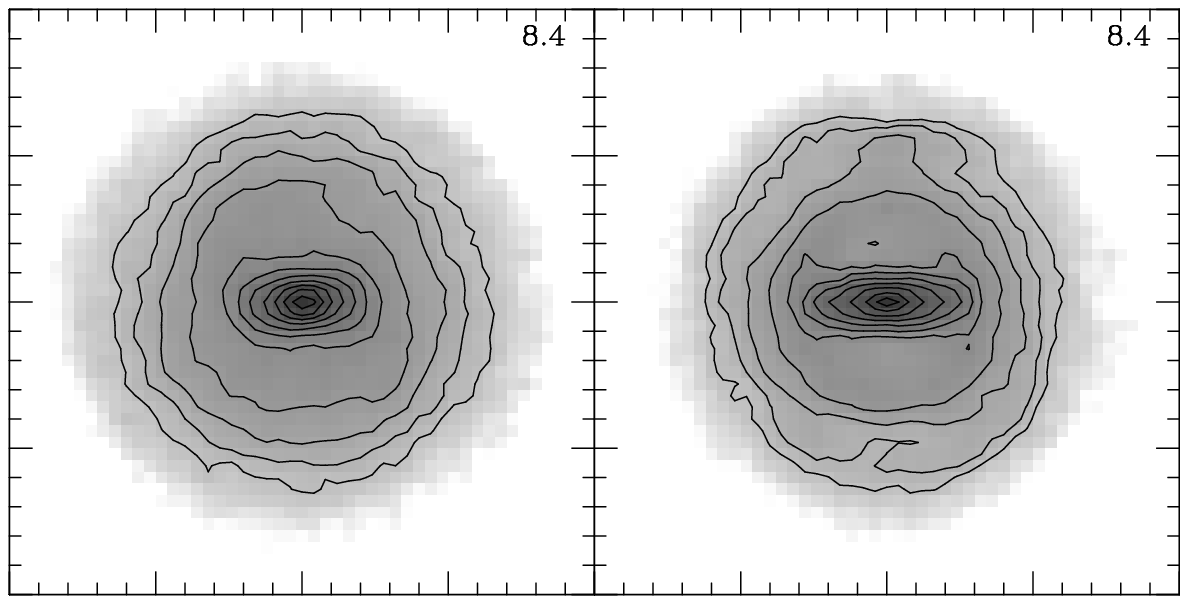}
\includegraphics[scale=0.76]{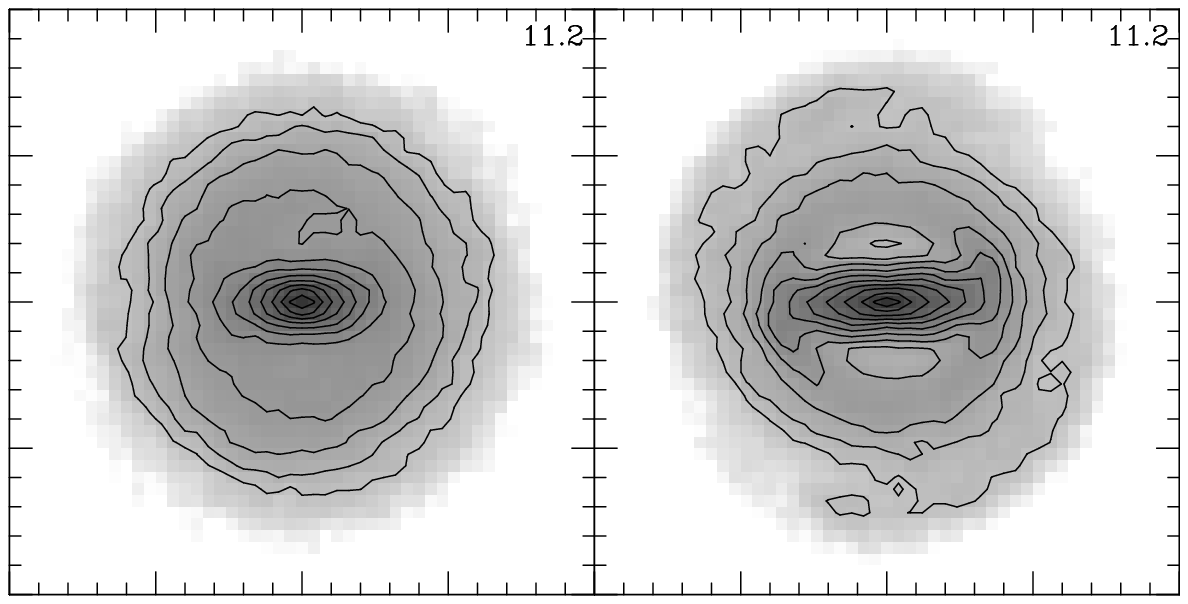}
\caption{Morphology of the disk component, viewed face-on, for an MD-type (left
panels) and an MH-type halo (right panels). The time in Gyr is given in the
upper right corner of each panel. }
\label{fig:MD-MH-morpho}
\end{center}
\end{figure*}
%--------------------------------------------------------------------

As shown in AM02 and illustrated in Fig.~\ref{fig:MD-MH-morpho}, the observable
properties of the bars which grow in these two types of models are quite
different. MH-type bars are stronger (longer, thinner and/or more massive) than
MD-type bars. Viewed face-on, they have a near-rectangular shape, while MD-type
bars are more elliptical. Viewed side-on, they show stronger peanuts and
sometimes (particularly towards the end of the simulation) even `X' shapes. On
the other hand, bars in MD-type models are predominantly boxy when viewed
side-on.

Thinking in terms of angular momentum exchange, it is easy to understand why
MH-type bars are stronger than MD-type ones. Indeed, the radial density profile
of MH-type haloes is such that, for reasonable values of the pattern speed, they
have more material at resonance than do MD- types. Thus, all else being similar,
there will be more angular momentum absorbed. This, in good agreement with
analytical results, should lead to stronger bars.

It should be stressed that the above discussion does not imply that all real
galaxies are either of MH type or of MD type. The two models illustrated here
were chosen as two examples, enclosing a useful range of halo radial density
profiles, which could actually be smaller than what is set by the two above
examples. Real galaxies can well be intermediate, i.e., somewhere in between the
two. It is nevertheless useful to describe the two extremes separately, since
this gives a better understanding of the effects of the spheroid. 

Models of MH- or MD-type which also have a classical bulge can be termed MHB and
MDB, respectively. The effect of the bulge in MD models is quite strong, so that
the bars in MDB models have a strength and properties which are intermediate
between those of MD and those of MH types (AM02). Furthermore, A03 and
Athanassoula (2007) showed that an initially non-rotating bulge absorbs a
considerable amount of angular momentum -- thereby spinning up -- and thus a bar
in a model with bulge slows down more than in a similar model but with no bulge.
All this can be easily understood from the frequency analysis, which shows that
there are considerably more particles at resonance in cases with strong bulges
(A03; Saha \textit{et al.} 2012). On the other hand, the effect of the bulge on
the bars of MH types is much less pronounced.    

\subsubsection{Models with cusps}
\label{subsubsec:cusps}

The two models we have discussed above have a core, more or less extended. There
is a further possibility, namely that the central part has a cusp. It has indeed
been widely debated whether haloes have a cusp or a core in their central parts.
Cosmological CDM, dark matter only simulations produce haloes with strong cusps.
Thus, Navarro \textit{et al.} (1996) find a universal halo profile, dubbed NFW
profile, which has a cusp with a central density slope ($\beta = d\ln\rho/d\ln
r$) of $-1.0$, while Moore \textit{et al.} (1999), with a higher resolution,
find a slope of $-1.5$. Increasing the resolution yet further, Navarro
\textit{et al.} (2004) found that this slope decreases with decreasing distance
from the centre, but not sufficiently to give a core. Finally, the simulations
with the highest resolution (Navarro \textit{et al.} 2010), argue for a lower
central slope of the order of $-0.7$, but still too high to be compatible with a
core. 

On the other hand, very extensive observational and modelling work (de Blok
\textit{et al.} 2001; de Blok \& Bosma 2002; de Blok \textit{et al.} 2003; Simon
\textit{et al.} 2003; Kuzio de Naray \textit{et al.} 2006, 2008; de Blok
\textit{et al.} 2008; Oh \textit{et al.} 2008; Battaglia \textit{et al.} 2008;
de Blok 2010; Walker \& Pe\~narrubia 2011; Amorisco \& Evans 2012; Pe\~narrubia
\textit{et al.} 2012) argues that the central parts of haloes should have a
core, or a very shallow cusp, the distribution of inner slopes in the various
observed samples of galaxies being strongly peaked around a value of
$\sim$\,0.2. This discrepancy between the pure dark matter, CDM simulations and
observations may be resolved with more recent cosmological simulations which
have high resolution and include baryons and appropriate star formation and
feedback recipes. Indeed, such simulations start to produce rotation curves
approaching those of observations (Governato \textit{et al.} 2010, 2012;
Macci\`o \textit{et al.} 2012; Oh \textit{et al.} 2011; Pontzen \& Governato
2012; Stinson \textit{et al.} 2012). In order to stay in agreement with
observations, I will here not discuss models with cusps. Readers interested in
such models can consult, e.g., Valenzuela \& Klypin (2003), Holley-Bockelmann
\textit{et al.} (2005), Sellwood \& Debattista (2006), or Dubinski \textit{et
al.} (2009).  Let me also mention that it is possible to study models with cusps
using the same functional form for the halo density as for the MH- and MD-type
models (AM02), but now taking a very small core radius, preferably of an extent
smaller than the softening length.

\subsubsection{The effect of the spheroid-to-disk mass ratio}
\label{subsubsec:MH-to-MD}

In the two models we discussed above, it is clear that it is the one with the
highest spheroid mass fraction within the disk region that makes the strongest
bar. Is that always the case? The following discussion, taken from A03, shows
that the answer is more complex than a simple yes, or no.

Assume we have a sequence of models, all with the same total mass, i.e., that
the sum of the disk and the spheroid mass within the disk region is the same.
How should we distribute the mass between the spheroid and the disk in order to
obtain the strongest bar? What must be maximised is the amount of angular
momentum redistribution, or, equivalently, the amount of angular momentum taken
from the bar region. For this it makes sense to have strong absorbers, who can
absorb all the angular momentum that the bar region can emit. Past a certain
limit, however, there will not be sufficient material in the bar region to emit
all the angular momentum that the spheroid can absorb, and it will be useless to
increase the spheroid mass further. So the strongest bar will not be obtained by
the most massive spheroid, but rather at a somewhat lower mass value, such that
the equilibrium between emitters and absorbers is optimum and the angular
momentum exchanged is maximum. For the models discussed in AM02 and A03, this
occurs at a spheroid mass value such that the disk, in the initial conditions,
is sub-maximum. In Section~\ref{subsec:max-submax} we will discuss how disks may
evolve from sub-maximum to maximum during the simulation.  
 
\subsection{Live versus rigid halo}
\label{subsec:rigid}

In the previous sections I reviewed the very strong evidence accumulated so far 
showing that many particles in the simulations, both in the disk and the
spheroid component, are on (near-)resonant orbits and that the angular momentum
exchanged between them is as predicted by the analytic calculations, i.e., from
the bar region outwards (Section~\ref{sec:analytic}). The next step should be to
clarify the importance of the halo resonances in the evolution. For this we have
to compare two simulations, one in which the halo resonances are at work and
another where they are not, as was first done in A02, whose main results will be
reviewed here.

\begin{figure*}[t]
\begin{center}
\includegraphics[scale=0.9]{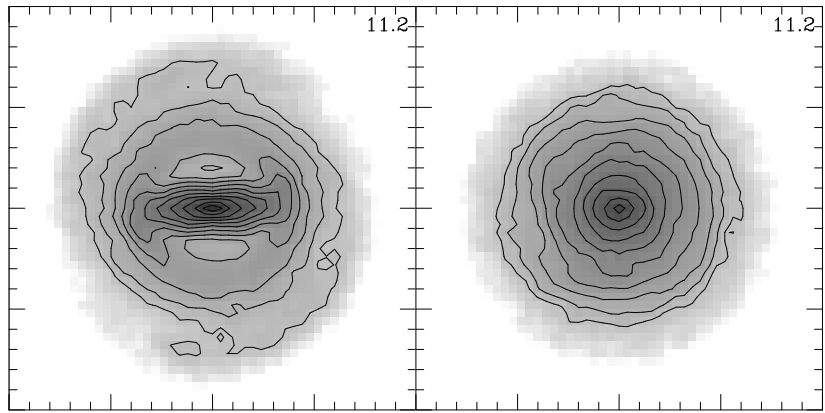}
\caption{Comparison of the evolution of the bar in a live halo (left-hand panel) 
to that in a rigid halo (right-hand panel) for an MH-type halo. }
\label{fig:MH_RH}
\end{center}
\end{figure*}

\begin{figure*}
\begin{center}
\includegraphics[scale=0.9]{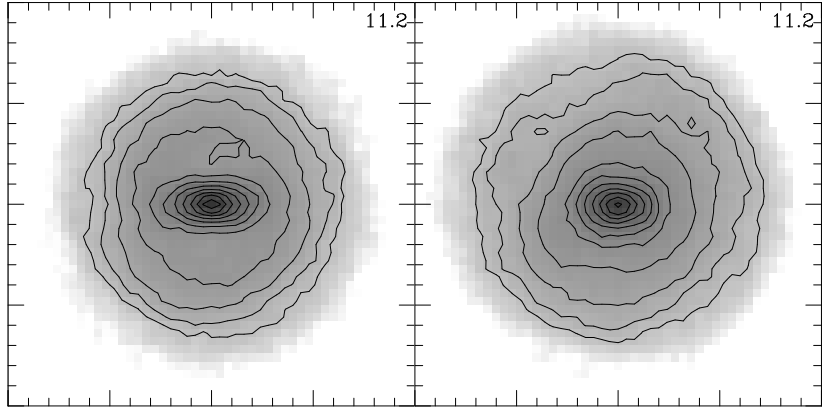}
\caption{Comparison of the evolution of the bar in a live halo (left-hand panel)
to that in a rigid halo (right-hand panel) for an MD-type halo.}
\label{fig:MD_RD}
\end{center}
\end{figure*}

Each of the Fig.~\ref{fig:MH_RH} and \ref{fig:MD_RD} compares two models with
initially identical disks. In other words, the particles in the disk initially
have identical positions and velocities in the two compared simulations. The
models of the haloes were also identical, but in one of the simulations
(right-hand panels) the halo was rigid (represented only by the forces that it
exerts on the disk particles) and thus did not evolve. In the other one,
however, the halo was represented by particles, i.e., was live (left-hand
panels). These particles move around as imposed by the forces and can emit or
absorb angular momentum, as required. 

%--------------------------------------------------------------------
%--------------------------------------------------------------------

Figure~\ref{fig:MH_RH} compares the disk evolution in the live and in the rigid
halo when the model is of MH type. The difference between the results of the two
simulations is stunning. In the case with a live halo a strong bar has formed,
while in the case with a rigid halo there is just a very small inner oval-like
perturbation. This shows that the contribution of the halo in the angular
momentum exchange can play an important role, actually, in the example shown
here, the preponderant role. 

%--------------------------------------------------------------------
%--------------------------------------------------------------------

Figure~\ref{fig:MD_RD}, shows the results of a similar experiment, but now in an
MD-type halo. The difference is not as stunning as in the previous example, but
is still quite important. In the live halo case the bar is considerably longer
and somewhat thinner than in the case with a rigid halo. 

It is thus possible to conclude that the role of the halo in the angular
momentum redistribution is important. In fact in the MH-type models the role of
the halo is preponderant, but it is still quite important even in the MD-types.
It is thus strongly advised to work with live, rather than with rigid haloes in
simulations.

\subsection{Distribution of frequencies for MD- and MH-type models}
\label{subsec:MHMDreson}

Figure~\ref{fig:resonances} displays the frequency histograms for two models,
one MD-type (upper panels) and one MH-type (lower panels). The properties of
these two types of models were discussed in Section~\ref{subsubsec:MH-MD}, where
their initial rotation curve, as well as their bar morphology are also
displayed.  
 
It is now useful to compare the distribution of frequencies for the two
simulations used in Fig.~\ref{fig:resonances}. Starting with the disk we note
that the ILR peak is about 50\% higher in the MH than in the MD model, while the
CR peak is considerably lower. Also the MD model has an OLR peak, albeit small,
while none can be seen in the MH one. For the spheroid, the strongest peak for
both models is the CR one, which is much stronger in the halo than in the disk.
It is, furthermore, stronger in the MH than in the MD model. Also the MH
spheroid has a relatively strong ILR peak, which is absent from the MD one. On
the other hand, the MD model has a much stronger OLR peak than the MH one.   

All these properties can be easily understood. From Fig.~\ref{fig:MD-MH-morpho}
it is clear that the bar in the MH model is stronger than in the MD one, as
discussed already in Section~\ref{subsubsec:MH-MD}, and this accounts for the
much stronger ILR peak for the disk of the MH model. Also, from the initial
circular velocity curves (Fig.~\ref{fig:rotcurves}) it is clear that the halo of
the MH model has much more mass than the MD model within the radial extent where
one would expect the CR and particularly the ILR to be. This explains why the CR
halo peaks are stronger in the MH model and why the halo ILR peak is absent in
the MD one. At larger radii the order between the masses of the MH and the MD
haloes is reversed, being in the outer parts relatively  larger in the MD model.
This explains why the halo OLR peak is stronger for the MD halo than for the MH
one.

\subsection{Bar strength}
\label{subsec:bar-str}

\subsubsection{Evolution of the bar strength with time}
\label{subsubsec:str_evol}

\begin{figure*}
\begin{center}
\includegraphics[scale=0.74]{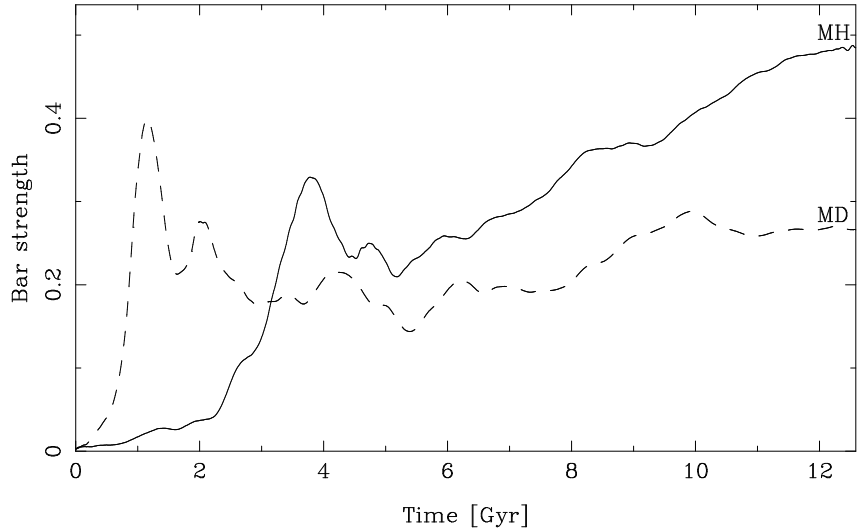}
\caption{Evolution of the bar strength with time for two models, one
of the MD-type (dashed line) and the other of the MH type (solid line).} 
\label{fig:str_evol}
\end{center}
\end{figure*}

Figure~\ref{fig:str_evol}  shows the evolution of the bar strength\footnote {The
definition of bar strength is not unique. The one used in the analysis of
simulations is usually based on the $m$\,=\,2 Fourier component, but precisely
how this is used varies from one study to another. We will refrain from giving a
list of precise definitions here, as this would be long and tedious. Furthermore
we will, anyway, only need qualitative information for our discussions here,
which is the same, or very similar, for all definitions used in simulations. We
will thus talk only loosely about `bar strength' here and use arbitrary units in
the plots (see also Section~\ref{sec:observations}).} with time, comparing an
MH-type and an MD-type model. It clearly illustrates how important the
differences between these two types can be, as expected. It also shows that, in
both cases, one can distinguish several evolutionary phases. 

%--------------------------------------------------------------------
%--------------------------------------------------------------------

By construction, both simulations start axisymmetric and this lasts all through
what we can call the pre-growth phase. The duration of this phase, however, is
about half a Gyr for the MD model, while for the MH one it lasts about 2\,Gyr.
The second phase is that of bar growth, and lasts considerably less than a Gyr
for the MD model and much longer (about 2\,Gyr) for the MH one. In total, we can
say that the bar takes less that 1\,Gyr to reach the end of its growth phase in
the MD model, compared to about 4\,Gyr in the MH one. This is in good
qualitative agreement with what was already found by Athanassoula \& Sellwood
(1986), using simpler 2D simulations. From this and many other such comparisons,
it becomes clear that the  presence of a massive spheroid can very considerably
both delay and slow down the initial bar formation due to its strong
contribution to the total gravitational force.  

After the end of the bar growth time, both models undergo a steep drop of the
bar strength. This is due to the buckling instability (Raha \textit{et al.}
1991). The final phase -- which can also be called the secular evolution phase
-- starts somewhat after 5\,Gyr for the MH model and after about 3\,Gyr for the
MD one. The corresponding bar strength increase which takes place during this
phase is much more important for the MH than for the MD mode. By the end of the
evolution, {\it MH models have a much stronger bar than MD ones}. As already
mentioned, this is due to the more important angular momentum redistribution in
the former type of models.

As in Section~\ref{subsubsec:MH-MD}, let me stress that we are comparing two
models which display strong differences. Real galaxies can be of either type,
but, most probably, can be intermediate, in which case their bar strength
evolution would also be intermediate between the two shown in
Fig.~\ref{fig:str_evol}.

\subsubsection{Spheroid mass and bar strength}
\label{subsubsec:MH-str}

Figure~\ref{fig:str_evol} illustrates another important point, first argued in
A02, namely that the effect of the spheroid on the bar strength is very
different, in fact opposite, in the early and late evolutionary stages. 

\begin{enumerate}[(a)]\listsize
\renewcommand{\theenumi}{(\alph{enumi})}

\item In the early evolutionary stages, before and while the bar grows, the
spheroid delays and slows down bar formation. This is due to the fact that the
gravitational forcing of the spheroid `dilutes' the non-axisymmetric forcing of
the bar. Thus, this delay and slowdown occurs even in cases with a rigid
spheroid, or with an insufficient number of particles (e.g., Ostriker \& Peebles
1973). 

\item In the late evolutionary stages, e.g., when the secular evolution is
underway, the presence of a massive and responsive spheroid will make the bar
much stronger. This is due to the help of the spheroid resonances, which absorb
a considerable fraction of the emitted angular momentum, thus inducing the bar
region to emit yet more and (since it is within the CR and of negative energy)
to become stronger. In order for this phase to be properly described the
spheroid has to be live and contain a sufficient number of particles for the
resonances to be properly described (A02). 

\end{enumerate}

%--------------------------------------------------------------------
\begin{figure*}
\begin{center}
\includegraphics[scale=0.74]{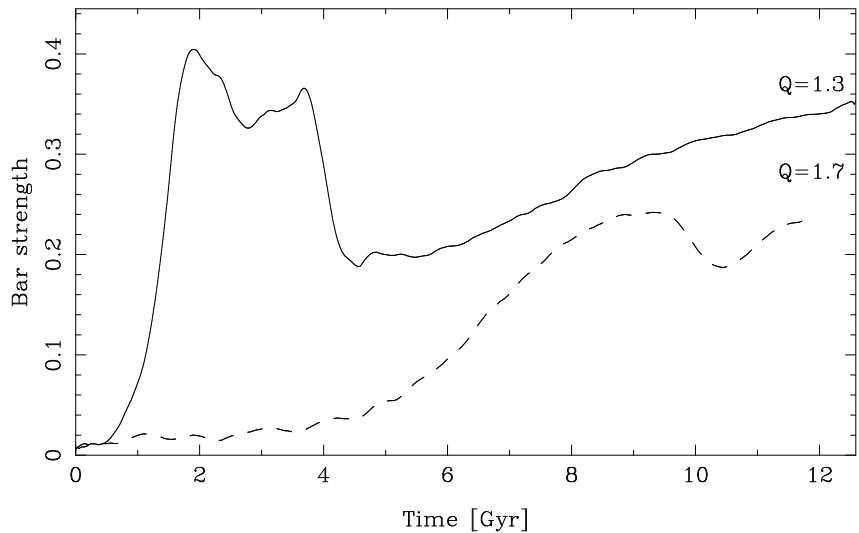}
\caption{Evolution of the bar strength with time for two models with different
Toomre $Q$ parameters. Both models are of MD type.} 
\label{fig:sigma-str}
\end{center}
\end{figure*}
%--------------------------------------------------------------------

\subsubsection{Velocity dispersion and bar strength}
\label{subsubsec:sigma-str}

Analytical works  for the disk and/or the spheroidal component (Lynden-Bell \&
Kalnajs 1972; Tremaine \& Weinberg 1984; A03) predict that the hotter the
(near-)resonant material is, the less angular momentum it can emit or absorb.
This was verified with the help of $N$-body simulations in A03. Contrary to the
spheroid mass, velocity dispersion has the same effect on the bar strength
evolution both in the early bar formation stages and in the later secular
evolution stages. In the early stages a high velocity dispersion in the disk
slows down bar formation, as shown initially by Athanassoula \& Sellwood (1986)
and later in A03. This is illustrated also in Fig.~\ref{fig:sigma-str}, where I
compute two MD-type models with different velocity dispersions. The first one
has a Toomre $Q$ parameter (Toomre 1964) of 1.3 and the second one of 1.7. This
difference has a considerable impact on the growth and evolution of the bar. In
the former the bar starts growing after roughly half a Gyr, its growth phase
lasts about 1.5\,Gyr and the secular increase of the bar strength starts around
4.5\,Gyr. For the latter (hotter) model the beginning and end of each phase are
much less clear, so that one can only very roughly say that the bar growth
starts at about 4, or 5\,Gyr and ends at about 9\,Gyr. During the later
evolutionary stages also, a high velocity dispersion will work against an
increasing bar strength because, as shown by analytic work and verified by
$N$-body simulations, material at resonance will emit or absorb per unit mass
less angular momentum when it is hot. 

Thus, {\it increasing the velocity dispersion in the disk and/or the spheroid
leads to a delayed and slower bar growth and to weaker bars}. This has important
repercussions on the fraction of disk galaxies that are barred as a function of
redshift and on their location on the Tully Fisher relation (Sheth \textit{et
al.} 2008, 2012). 

\subsubsection{Bar strength and redistribution of angular momentum}  
\label{subsubsec:correl}

\begin{figure}
\begin{center}
\includegraphics[scale=0.8]{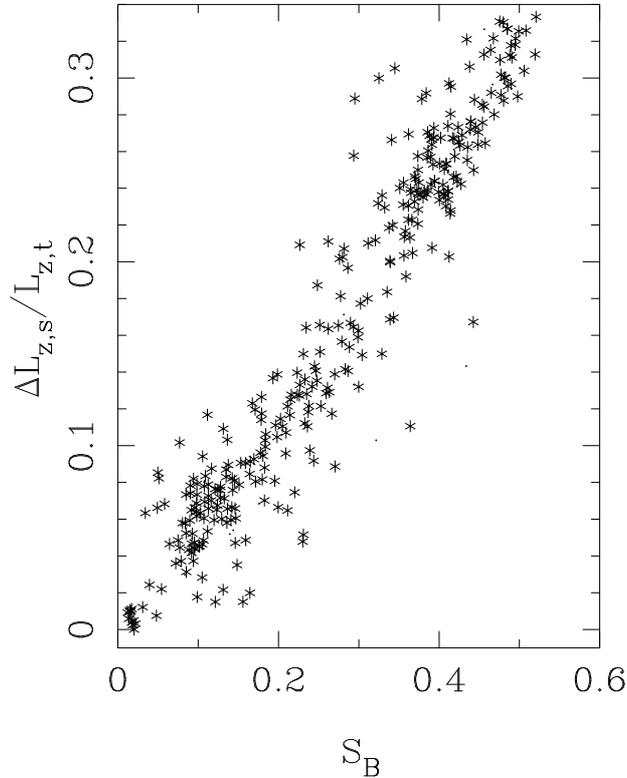}
\caption{Bar strength, $S_{\rm B}$, as a function of the amount of angular
momentum absorbed by the spheroid, $\Delta L_{z,{\rm s}}$, expressed as a
fraction of the total $z$ component of the angular momentum, $L_{z,{\rm t}}$.
The two quantities are measured at the same time, towards the end of the
simulation when all bars are in their secular evolution phase. Each symbol in
this plot represents a separate simulation.} 
\label{fig:str_jmom_correl}
\end{center}
\end{figure}

One of the predictions of the analytic work is that there is a strong link
between the angular momentum which is redistributed within the galaxy and the
bar strength.  One may thus expect a correlation between the two if the
distribution functions of the disks and spheroids of the various models are not
too dissimilar. This was tested out in A03, using a total of 125 simulations,
and was found to be true. Here we repeat this test, using a somewhat larger
number of simulations (about 400 instead of 125) and a more diverse set of
models and again a good correlation is found. The result is shown in
Fig.~\ref{fig:str_jmom_correl}, where each symbol represents a separate
simulation. It is clear that this correlation is tight, but still has some
spread, due to the diversity of the models used. Note also that we have not
actually used the total amount of angular momentum emitted from the bar region
(or, equivalently, the amount of angular momentum absorbed in the outer disk
{\it and} in the spheroid), but rather the fraction of the total initial angular
momentum that was deposited in the spheroid, which proves to be a good proxy to
the required quantity. Finally, note that the points are not homogeneously
distributed along the trend. This has no physical significance, but is simply
due to the way that I chose my simulations. Indeed I tried to study the MH-type
and the MD-type models and was relatively less interested in the intermediate
cases. 

%--------------------------------------------------------------------
%--------------------------------------------------------------------

%
%%%%%%%%%%%%%%%%%%%%%%%%%%%%%%%%%%%%%%%%%%%%%%%%%%%%%%%%%%%%%%%%%%%%%%%%%%%%%%%%
%

\section{Bar slowdown}
\label{sec:slowdown}
 
\subsection{Results from N-body simulations}
\label{subsec:Nslowdown}

Another prediction of the analytic work is that the bar pattern speed will
decrease with time, as the bar strength increases (Section~\ref{sec:analytic}).
This has been confirmed by a large number of $N$-body simulations (e.g.,
Little \& Carlberg 1991a,\,b; Hernquist \& Weinberg 1992; Debattista \& Sellwood
2000; A03; O'Neil \& Dubinski 2003; Mart\'inez-Valpuesta \textit{et al.} 2006;
Villa-Vargas \textit{et al.} 2009). The amount of this decrease was found to
vary considerably from one simulation to another, depending on the mass as well
as on the velocity distribution in the disk and the spheroidal (halo plus
classical bulge) components, consistent with the fact that these mass and
velocity distributions will condition the angular momentum exchange and
therefore the bar slowdown.

There is a notable exception to the above very consistent picture. Valenzuela \&
Klypin (2003) found in their simulations a counter-example to the above, where
the pattern speed of a strong bar hardly decreases over a considerable period of
time. The code they use, ART, includes adaptive mesh refinement, and thus
reaches high resolution in regions where the particle density is high. According
to these authors, the difference between their results and those of other
simulations are due to the high resolution (20--40\,pc) and the large number of
particles (up to 10$^7$) they use. Sellwood \& Debattista (2006) examined cases
where the bar pattern speed fluctuates upward. After such a fluctuation, the
density of resonant halo particles will have a local inflection created by the
earlier exchanges, so that bar slowdown can be delayed for some period of time.
They show that this is more likely to occur in simulations using an adaptive
refinement and propose that this explains the evolution of the pattern speed in
the simulation of Valenzuela \& Klypin (2003). Klypin \textit{et al.} (2009) did
not agree and replied that Sellwood \& Debattista did not have the same adaptive
refinement implementation as ART. Sellwood (2010) stressed that such episodes
of non-decreasing pattern speed are disturbed by perturbations, as e.g., a halo
substructure, and thus are necessarily short lived. He thus concludes that
simulations where the pattern speed does not decrease have simply not been run
long enough. At the other extreme, Villa-Vargas {\it et al.} (2009) find a
similar stalling of the pattern speed for prolonged time periods when the
simulation is run so long that the corotation radius gets beyond the edge of the
disc.

Dubinski \textit{et al.} (2009) published a series of simulations, all with the
same model but with increasing resolution. They use between 1.8$\times$10$^3$ 
and 18$\times$10$^6$ particles in the disk and between 10$^4$ and 10$^8$
particles in the halo. They also present results from a multi-mass model with an
effective resolution of $\sim$\,10$^{10}$ particles. They have variable, density
dependent softening, with a minimum of the order of 10\,pc. Their Fig.~18 shows
clearly that the decrease of the pattern speed with time does not depend on the
resolution and that it is present for all of their simulations, even the ones
with the highest resolution, much higher than the one used by Valenzuela \&
Klypin. They conclude that `the bar displays a convergent behavior for halo
particle numbers between 10$^6$ and 10$^7$ particles, when comparing bar growth,
pattern speed evolution, the DM [dark matter] halo density profile and a
nonlinear analysis of the orbital resonances'. This makes it clear that, at
least for their model, the pattern speed decreases with time for all
reasonable values of particle numbers. 

%--------------------------------------------------------------------
\begin{figure*}
\begin{center}
\includegraphics[scale=0.74, angle=-90]{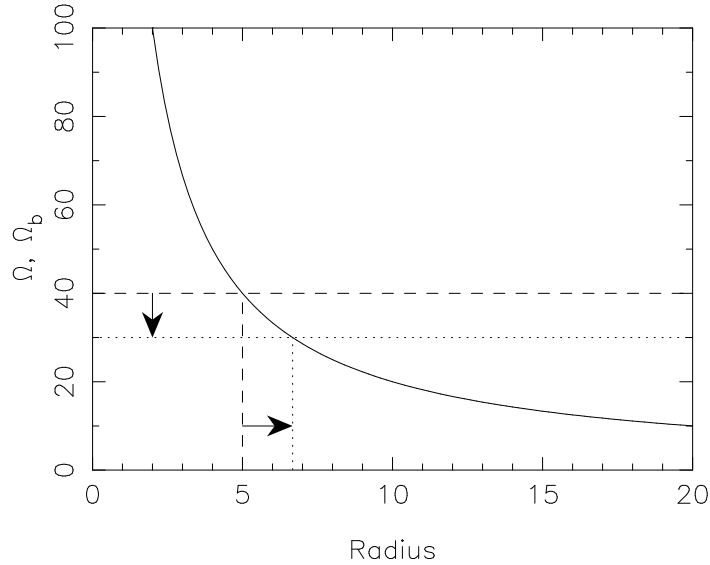}
\caption{Schematic plot of the effects of the decrease of the bar pattern speed
with time for a very simple model with a constant circular velocity. The solid
line gives $\Omega (r)$ and the horizontal dashed and dotted lines give the
pattern speed $\Omega_{\rm p}$ at two instants of time. The vertical dashed and
dotted lines show the CR radius at these same two instants of time. The vertical
arrow indicates the decrease of the bar pattern speed and the horizontal one the
increase of the corotation radius.} 
\label{fig:schema_evol}
\end{center}
\end{figure*}
%--------------------------------------------------------------------
 
\subsection{A schematic view}
\label{subsec:Nschema}

Figure~\ref{fig:schema_evol} shows very schematically an interesting effect of
the bar slowdown. The solid line shows the radial profile of $\Omega(r)$ for a
very simple model with a constant circular velocity, but the following hold for
any realistic circular velocity curve. Let us assume that at time $t$\,=\,$t_1$
the pattern speed is given by the dashed horizontal line and drops by
$t$\,=\,$t_2$ to a lower value given by the dotted horizontal line, as shown by
the vertical arrow. This induces a change in the location of the resonances. For
example the CR, which at $t_1$ is located at 5\,kpc, as given by vertical dashed
line, will move by $t_2$ considerably outwards to a distance beyond 6\,kpc, as
given by the dotted vertical line and shown by the horizontal arrow. This
increases also the region in which the bar is allowed to extend, since, as shown
by orbital theory (Contopoulos 1980 and Section~\ref{sec:orbits}), the bar size
is limited by the CR. This schematic plot also makes it easy to understand how a
`fast' bar can slow down considerably while remaining `fast'. As we saw in
Section~\ref{sec:orbits}, a bar is defined to be `fast' if the ratio
$\mathcal{R}$ of the corotation radius to the bar length is less than 1.4. Thus,
a bar that slows down and whose corotation radius\linebreak increases can still have
$\mathcal{R}$\,$<$\,1.4, provided the bar length increases accordingly. This
occurs in a number of simulations, see, e.g., Dubinski \textit{et al.} (2009). 
 
\subsection{What sets the pattern speed value?}
\label{subsec:set-omegap}

What sets the value of the pattern speed in a simulation (and thus also
presumably in real galaxies)? The value of the pattern speed is set by the value
of the corotation radius, which is in fact the borderline between emitters and
absorbers. Thus, if the galaxy wants to maximise the amount of angular momentum
it pushes outwards (i.e., the amount of angular momentum that it
redistributes), it has to set this boundary, and therefore its bar pattern
speed, appropriately. If the spheroid is massive, i.e., if it has sufficient
mass in the resonant regions, then the bar can lower its pattern speed in order
to have more emitters, since the absorbers are anyway strong. On the other hand
if the spheroid is not sufficiently massive, then the bar should not lower its
pattern speed overly, because it needs the absorption it can get from the outer
disk. Thus, indirectly, it could be the capacity of the spheroid resonances to
absorb angular momentum that sets the value of the bar pattern speed. This would
mean that properties of the dark matter halo and of the classical bulge, such as
their mass relative to that of the disk and their velocity dispersion at the
resonant regions, will have a crucial role in setting the bar pattern speed.

%
%%%%%%%%%%%%%%%%%%%%%%%%%%%%%%%%%%%%%%%%%%%%%%%%%%%%%%%%%%%%%%%%%%%%%%%%%%%%%%%%
%

\section{Boxy/peanut bulges}
\label{sec:BP}
 
\subsection{Peanuts: input from simulations, orbits and observations}
\label{subsec:BPgeneral}

When bars form in $N$-body simulations they have a thin vertical density
profile, similar to that of the disk. In other words, it is the in-plane
rearrangement of the disk material that creates the bar, when initially
near-planar and near-circular orbits become more elongated and material gets
trapped around the stable periodic orbits of the $x_1$ family, as already
discussed in Section~\ref{sec:orbits} and \ref{subsec:angmom}. This
configuration, however, lasts for only a short while, after which the bar
buckles out of the plane and becomes asymmetric with respect to the equatorial
plane, as shown, e.g., in Combes \textit{et al.} (1990) and Raha \textit{et al.}
(1991), and as is illustrated in the middle central panel of
Fig.~\ref{fig:snapshot_evol}. This evolutionary phase, which can be called the
asymmetry phase, is also very short-lived and soon the side-on view displays a
clear peanut or boxy shape. During the peanut formation phase the strength of
the bar decreases considerably (Combes \textit{et al.} 1990; Raha \textit{et
al.} 1991; Debattista \textit{et al.} 2004, 2006; Mart\'inez-Valpuesta \&
Shlosman 2004; Mart\'inez-Valpuesta \textit{et al.} 2006; Athanassoula 2008a).
Two examples of this decrease can be seen in Fig.~\ref{fig:str_evol}, one for an
MH-type simulation (where the bar strength decrease starts only at roughly
6\,Gyrs) and another for an MD-type simulation (where it already starts at
roughly 3\,Gyrs). This decrease can sometimes be very important, so that it
could get erroneously interpreted as a bar destruction.   

These boxy/peanut structures had been observed in real galaxies many times, well
before being seen in simulations. Due to the fact that they extend vertically
well outside the disk, they were called bulges. More specifically, if they have
a rectangular-like (box-like) outline they are called boxy bulges, and if their
outline is more reminiscent of a peanut, they are called peanut bulges.
Sometimes, however, this distinction is not made and the words `boxes' or
`peanuts' are used indiscriminately, or the more generic term `boxy/peanut' is
used instead. A number of kinematical or photometrical observations followed and
comparisons of their results with orbits and with simulations established the
link of boxy/peanut bulges to bars (Kuijken \& Merrifield 1995; Athanassoula \&
Bureau 1999; Bureau \& Athanassoula 1999; Bureau \& Freeman 1999; Merrifield \&
Kuijken 1999; L\"utticke \textit{et al.} 2000; Aronica \textit{et al.} 2003;
Chung \& Bureau 2004; Athanassoula 2005b; Bureau \& Athanassoula 2005; Bureau
\textit{et al.} 2006).

\subsection{Peanut-related orbital structure}
\label{subsec:BPorbits}

Considerable information on boxy/peanut structures can be obtained with the help
of orbital structure theory. In 3D the orbital structure is much more complex
than in 2D, as expected. Thus, the $x_1$ family has many sections (i.e., energy
ranges) where its members are vertically unstable, and, at the  energies where
there is a transition from stability to instability, a 3D family can bifurcate
(i.e., emerge). The orbits that are trapped around the stable $l$\,=\,1,
$m$\,=\,2, $n$\,$\neq$\,0 periodic orbits of this family can participate in the
boxy/peanut structure (Patsis \textit{et al.} 2003). They were discussed by
Pfenniger (1984) and by Skokos \textit{et al.} (2002a,\,b), who presented and
described a number of relevant families. Since these orbits  bifurcate from the
$x_1$ and create vertically extended structures, they were named by Skokos
\textit{et al.} (2002a) by adding a $v_i$,$i$\,=\,1,\,2,\,... after the $x_1$,
i.e., $x_1v_i$, $i$\,=\,1,\,2,\,..., where $i$ is the order of the bifurcation.
Projected on the $(x,y)$ plane, their shape is very  similar to that of the
members of the planar $x_1$ family. Good examples of such periodic orbits can be
seen in Fig.~9 of Pfenniger (1984), or Fig.~7 to 10 of Skokos \textit{et al.}
(2002a). 

\subsection{Peanuts as parts of bars: shape and extent}
\label{subsec:BPextent}

Contrary to what has been very often said and written, boxy/peanut bulges are
{\it not} bars seen edge-on. The correct statement is that boxy/peanut bulges
are {\it the inner parts of} bars seen edge-on. The evidence for this was put
together and discussed in Athanassoula (2005b) and I will only summarise it
briefly here. Orbital structure theory shows that not all planar periodic orbits
of the $x_1$ family are vertically unstable. In fact, the ones in the outer part
of the bar are stable. Therefore, the outer part of the bar will stay thin and
only the part within a given radius will thicken, so that the peanut will be
shorter than the bar. This gives the bar an interesting form. As a very rough
approximation, one can think of the bar as a rectangular parallelepiped box
(like a shoe box), from the two smallest sides of which (perpendicular to the
bar major axis) stick out thin extensions. Of course this is a very rough
picture and the shape of the `box' is in fact much more complex than a
rectangular parallelepiped, while the extensions have shapes which are difficult
to describe. The best is to look at an animation\footnote{{\tt http://lam.oamp.fr/recherche-14/dynamique-des-galaxies/scientific-results/milky-way/\linebreak bar-bulge/how-many-bars-in-mw}, or in 
{\tt http://195.221.212.246:4780/dynam/movie/MWbar}.} where one can see a bar from a
simulation, from various viewing angles.

How much longer is the bar than the peanut? The answer to this question is not
unique and depends on which one of the $x_1v_i$ families sets the end of the
peanut, on the galactic potential and on the bar pattern speed (Patsis
\textit{et al.} 2003). 

%--------------------------------------------------------------------
\begin{figure*}
\begin{center}
\includegraphics[scale=0.9, angle=0]{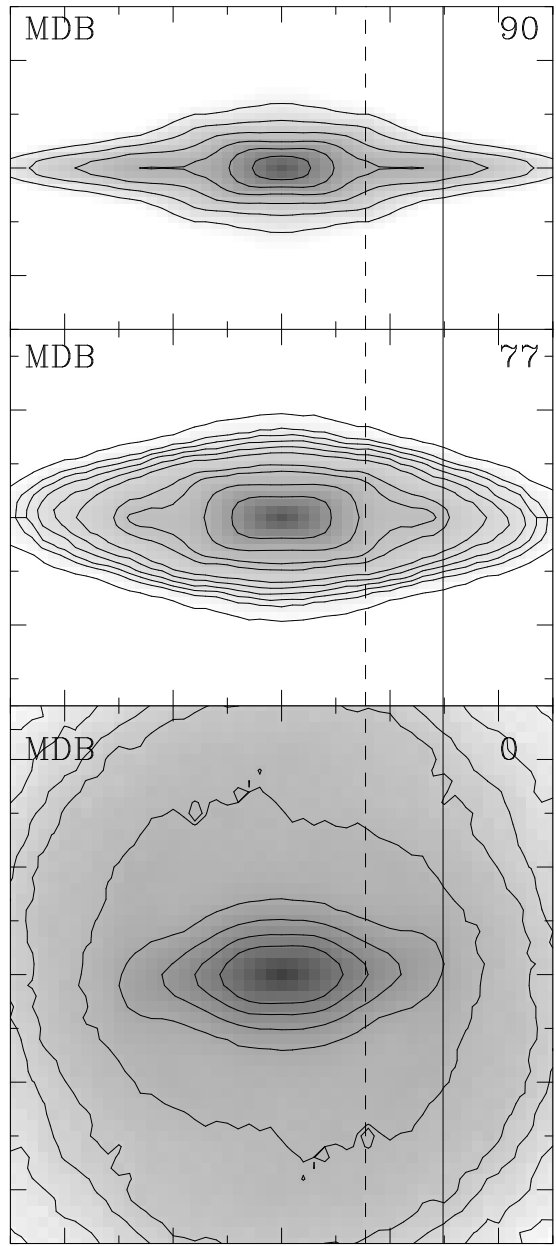}
\caption{Three views of the baryonic components (disk and bulge) of a
simulation. For each panel the inclination angle is given in the upper right
corner, the value of 77 degrees corresponding to the inclination of the
Andromeda galaxy (M\,31). The solid vertical line gives an estimate of the bar
length from the face-on view, while the dashed vertical line gives an estimate
of the length of the boxy/peanut structure, as obtained from the side-on view.} 
\label{fig:bp-length-th}
\end{center}
\end{figure*}
%--------------------------------------------------------------------

Figure~\ref{fig:bp-length-th} gives an estimate of the ratio of boxy/peanut to
thin bar length, for one of my simulations. In general, it is much easier to
obtain an estimate of this ratio for simulations than for observed galaxies,
because one can view snapshots from any desired angle. Thus, the length of the
bar can be obtained from the face-on view (lower panel) as the major axis of the
largest isophotal contour that has a bar shape. This of course introduces an
uncertainty of a few to several percent, but is about as good as one can achieve
with difficult quantities such as the bar length\footnote[2]{A discussion of the
various methods that can be used to measure a bar length in simulations, in
ensembles of orbits and/or in observations, and of their respective advantages
and disadvantages can be be found in AM02, Patsis \textit{et al.} (2002),
Michel-Dansac \& Wozniak (2006) and Gadotti \textit{et al.} (2007).}. The size
of the boxy/peanut part can be found from the edge-on view (upper panel). This
also introduces an uncertainty, probably much larger than that of the bar length
(AM02), but even so one can get reasonable estimates of the ratio of the two
extents, and certainly make clear that the thin part of the bar can be much
longer than the thick boxy/peanut part. Further discussion of this, and further
examples can be found in Athanassoula (2005b), Athanassoula \& Beaton (2006) and
Athanassoula (2008a).

\begin{figure*}
\begin{center}
\includegraphics[scale=0.57, angle=0]{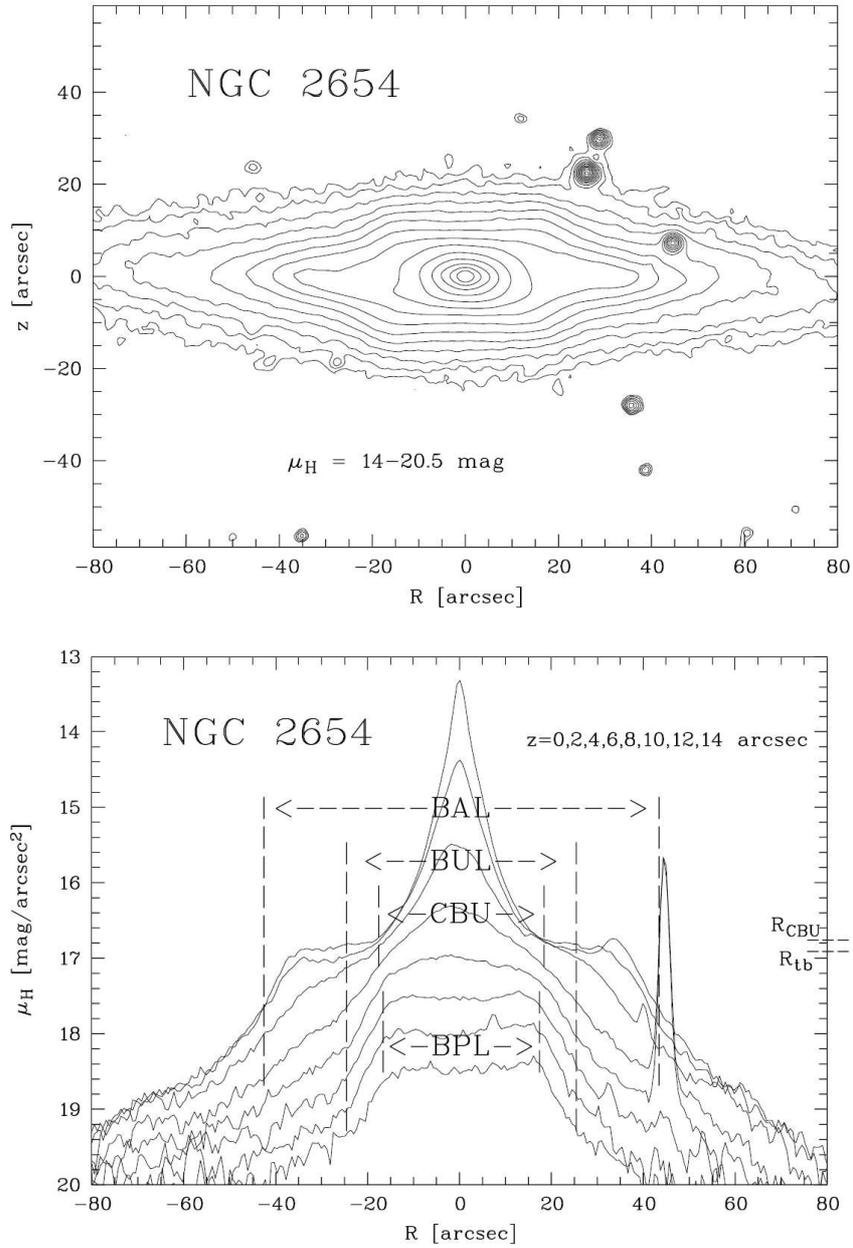}
\caption{Upper panel: Isophotes for the edge-on disk galaxy NGC\,2654 in the
near-infrared. Lower panel: Surface brightness profiles from cuts along, or
parallel to the major axis of this edge-on disk galaxy. From the cut along the
major axis (uppermost curve), it is possible to obtain an estimate of the
projected bar length -- BAL on the plot. The size of the boxy/peanut bulge is
obtained from cuts parallel to the major axis, but offset from it above or below
the equatorial plane -- BPL on the plot. (Figure~1 of L\"utticke \textit{et al.}
2000, reproduced with permission \copyright~ESO).} 
\label{fig:bp-length-obs}
\end{center}
\end{figure*}

Estimates of the ratio of boxy/peanut to thin bar length can also be obtained
from observations. Nevertheless, information for observed galaxies can be
obtained from only one viewing angle and these estimates are less precise than
the corresponding simulation ones. Figure~\ref{fig:bp-length-obs} allows us to
get an estimate for NGC\,2654. L\"utticke \textit{et al.} (2000) made a cut
along the major axis of this edge-on disk galaxy and from the projected surface
density profile along it they obtained the thin bar length (confront with method
(vi) from AM02). They also made cuts parallel to this and offset above or below
it and from them could obtain the extent of the bar/peanut part. In this way,
L\"utticke \textit{et al.} (2000) were able to measure the ratio of extent of
the thin part of the bar to the extent of the thick boxy/peanut part and
show clearly that the former can be much longer than the latter. 

%--------------------------------------------------------------------
%--------------------------------------------------------------------

\subsection{The boxy/peanut system in the Milky Way}
\label{subsec:milky-way}
 
The bar shape described in the previous section has important implications for
the structure of the Milky Way. It is now well established that our Galaxy is
barred (e.g., de Vaucouleurs 1964; Binney \textit{et al.} 1991; Blitz \& Spergel
1991). The thick component which can be seen to extend outside the Galactic
plane in the near-infrared \textit{COBE} (COsmic Background Explorer) image is
often referred to as the \textit{COBE}/DIRBE (Diffuse Infrared Background
Experiment) bar, or the thick bar. About ten years later, further evidence
started accumulating and was initially interpreted as due to the existence of a
second bar, longer than the first one and considerably thinner (Hammersley
\textit{et al.} 2000; Benjamin \textit{et al.} 2005; Cabrera-Lavers \textit{et
al.} 2007; L\'opez-Corredoira \textit{et al.} 2007; Cabrera-Lavers \textit{et
al.} 2008; Churchwell \textit{et al.} 2009). This second bar has been named the Long bar.
The existence of a second bar is very common in barred galaxies and about a
fourth or a fifth of disk galaxies have both a primary or main bar and a
secondary or inner bar (Erwin \& Sparke 2002; Laine \textit{et al.} 2002; Erwin
2011). However, the ratio of the lengths of the two presumed Milky Way bars is
totally incompatible with what is observed in double-barred external galaxies
(Romero-G\'omez \textit{et al.} 2011), and it would be very dangerous to assume
that our Galaxy has morphological characteristics so different from those of
external galaxies. 

There are two very important clues that can help us understand the nature of the
bar system in the Milky Way. The first one is that the \textit{COBE}/DIRBE bar
is thick while the Long bar is thin, their ratio of minor ($z$-) axis to major
axis being of the order of 0.3 and 0.03, respectively. The second one is that
the \textit{COBE}/DIRBE bar is shorter than the Long bar by a factor of roughly
0.8. These clues, taken together with the discussion in
Sections~\ref{subsec:BPorbits} and \ref{subsec:BPextent}, point clearly to a
solution where the thick \textit{COBE}/DIRBE bar and the thin bar are just parts
of the same single bar, the former being its thick boxy/peanut part and the
latter being its outer thin part. This alternative was first proposed for our
Galaxy by Athanassoula (2006, 2008b) and first tested by Cabrera-Lavers
\textit{et al.} (2007) using their red-clump giant measurements. This suggestion
was disputed at the time, because a number of observations (Hammersley
\textit{et al.} 2000; Benjamin \textit{et al.} 2005; Cabrera-Lavers \textit{et
al.} 2008) argued that the position angles of the \textit{COBE}/DIRBE bar and of
the Long bar are considerably different, with values between 15 and 30 degrees
for the former and around 43 degrees for the latter.  

Yet this difference in orientations is not a very strong argument. First, due to
our location within the Galaxy, the estimates for the Galactic bar position
angles are much less accurate than the corresponding estimates for external
galaxies. Thus, Zasowski \textit{et al.} (2012) find the position angle of the
Long bar to be around 35 degrees, i.e., much closer to that of the
\textit{COBE}/DIRBE bar than the 43 degrees estimated in previous works. Second,
if the shape of the outer isodensity contours of the thin part of the bar are,
in the equatorial plane, more rectangular-like than elliptical-like -- as is
often the case in external galaxies (e.g., Athanassoula \textit{et al.} 1990;
Gadotti 2008, 2011) -- the Long bar position angle will appear to be larger than
what it actually is. A third is that our Galaxy could well have an inner ring,
of the size of the bar. $N$-body simulations have shown that, in such cases,
there is often within the ring a short, leading segment near the end of the bar.
Examples can be found in Fig.~2 in AM02, Fig.~3 and 12 in A03, Fig.~2 in
Romero-G\'omez \textit{et al.} (2011), or Fig.~1 in Mart\'inez-Valpuesta \&
Gerhard (2011). Such a segment can spuriously increase the observed position
angle of the Long bar.

In view of all the above comments, the small difference between the position
angle of the \textit{COBE}/DIRBE bar and that of the Long bar should not be a
major concern. I thus still believe that my initial proposal, that the
\textit{COBE}/DIRBE and the Long bar are parts of the same bar, is correct.

%
%%%%%%%%%%%%%%%%%%%%%%%%%%%%%%%%%%%%%%%%%%%%%%%%%%%%%%%%%%%%%%%%%%%%%%%%%%%%%%%%
%

\section{Secular evolution of the disk and of its substructures}
\label{sec:disk}
 
The presence of a bar induces not only the redistribution of angular momentum
within the host galaxy (Section~\ref{sec:analytic} and \ref{sec:angmom}), but also
the redistribution of  the  material within it. The torques it exerts are such
that material within the CR is pushed inwards, while material outside the CR is
pushed outwards. As a result, there is a considerable redistribution of the
disk mass. 

\subsection{Redistribution of the disk mass:\\ formation of the disky bulge}
\label{subsec:discy-bulge}

It is well known that gas will concentrate to the inner parts of the disk under
the influence of the gravitational torque of a bar, thus forming an inner disk
whose extent is of the order of a kpc (Athanassoula 1992b; Wada \& Habe 1992,
1995; Friedli \& Benz 1993; Heller \& Shlosman 1994; Sakamoto \textit{et al.}
1999; Sheth \textit{et al.} 2003; Regan \& Teuben 2004). When this gaseous disk
becomes sufficiently massive it will form stars, which should be observable as a
young population in the central part of disks. Kormendy \& Kennicutt (2004)
estimate that the star formation rate density in this region is
0.1--1\,$M_\odot$\,yr$^{-1}$\,kpc$^{-2}$, i.e., one to three orders of magnitude
higher than the star formation rate averaged over the whole disk. Such disks can
harbour a number of substructures, such as spirals, rings, bright star-forming
knots, dust lanes and even (inner) bars, as discussed, e.g., in Kormendy (1993),
Carollo \textit{et al.} (1998) and Kormendy \& Kennicutt (2004). Furthermore, a
considerable amount of old stars is pushed inwards so that this inner disk will
also contain a considerable fraction of old stars (Grosb{\o}l \textit{et al.}
2004). Such disks are thus formed in $N$-body simulations even when the models
have no gas, as seen, e.g., in AM02, or Athanassoula (2005b). 

Such inner disks are evident in projected surface luminosity radial profiles, as
extra light in the central part of the disk, above the exponential profile
fitting the remaining (non-central) part. Since this is one of the definitions
for bulges, such inner disks have been linked to bulges. When fitting them with
an $r^{1/n}$ law -- commonly known as S\'ersic's law (S\'ersic 1968) -- the
values found for $n$ are of the order of or less than 2 (Kormendy \& Kennicutt
2004 and references therein). They are thus often called disky bulges, or
disk-like bulges (Athanassoula 2005b), or pseudobulges (Kormendy 1993, Kormendy
\& Kennicutt 2004).

\subsection{Redistribution of the disk mass:\\ the disk scale-length and extent}

Due to the bar torques and the resulting mass redistribution, the parts of the
disk beyond corotation become more extended and the disk scale length increases
considerably (e.g., Hohl 1971; Athanassoula \& Misiriotis 2002; O'Neil \&
Dubinski 2003; Valenzuela \& Klypin 2003; Debattista \textit{et al.} 2006;
Minchev \textit{et al.} 2011). Debattista \textit{et al.} 2006 showed that the
value of Toomre $Q$ parameter (Toomre 1964) of the disk can strongly influence how
much this increase will be.

Important extensions of the disk can also be brought about by flux-tube manifold
spiral arms (Romero-G\'omez \textit{et al.} 2006, 2007; Athanassoula \textit{et al.}
2009a, 2009b, 2010), as shown by Athanassoula (2012) who reported a strong
extension of the disk size, by as much as 50\% after two or three episodes of
spiral arm formation within a couple of Gyrs. 

\subsection{Redistribution of the disk mass:\\ maximum versus sub-maximum disks}
\label{subsec:max-submax}
 
Sackett (1997) and Bosma (2000) discuss a simple, straightforward criterion
allowing us to distinguish maximum from sub-maximum disks. Consider the ratio $S
= V_{\rm d,max} / V_{\rm tot}$, where $V_{\rm d,max}$ is the circular velocity
due to the disk component and $V_{\rm tot}$ is the total circular velocity, both
calculated at a radius equal to 2.2 disk scalelengths. According to Sackett
(1997), this ratio has to be at least 0.75 for the disk to be considered
maximum. Of course in the case of strongly  barred galaxies the velocity field
is non-axisymmetric and one should consider azimuthally averaged rotation
curves, or `circular velocity' curves. Furthermore, in the case of strongly
barred galaxies it is not easy to define a disk scalelength, so it is better to
calculate $S$ at the radius at which the disk  rotation curve is maximum, which
is a well-defined radius and is roughly equal to 2.2 disk scalelengths in the
case of an axisymmetric exponential disk. After these small adjustments, we can
apply this criterion to our simulations. 

\begin{figure}
\begin{center}
\includegraphics[scale=0.50, angle=0]{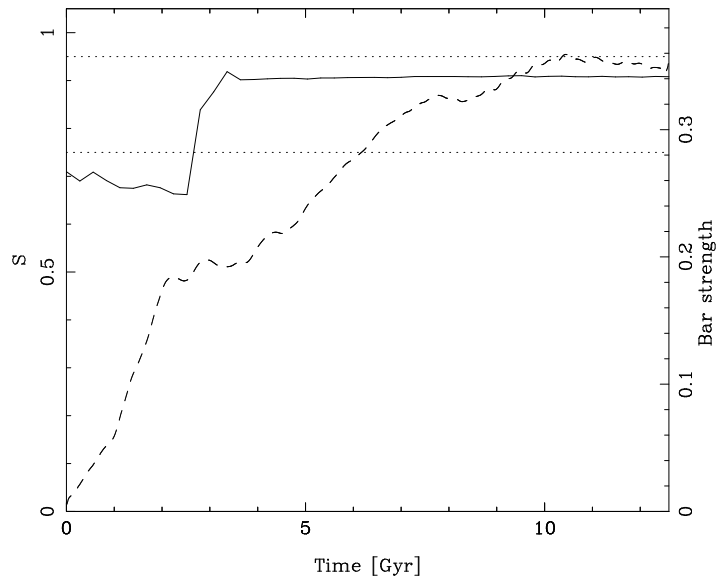}
\caption{Evolution of the Sackett parameter, $S$, as a function of time for an
MH-type simulation (solid line). The two horizontal dotted lines give the limits
within which $S$ must lie for the disk to be considered as maximum. The dashed
line gives a measure of the bar strength as a function of time, for the same
simulation.} 
\label{fig:max-submax}
\end{center}
\end{figure}

In Section~\ref{subsubsec:MH-MD} we saw that the disks in MH models are
sub-maximum in the beginning of the simulation and in
Section~\ref{subsec:discy-bulge} that the bar can redistribute the disk
material and in particular push material inwards and create a disky bulge. Is
this redistribution sufficient to change sub-maximum disks? The answer is that
this can indeed be true in some cases, as was shown in Athanassoula (2002b) and
is illustrated in Fig.~\ref{fig:max-submax}. This shows the Sackett parameter
$S$ and the bar strength as a function of time for one such simulation. Note
that the disk is initially sub-maximum and that it stays so during the bar
growth phase. Then the value of $S$ increases very abruptly to a value larger
than 0.75, so that the disk becomes maximum. After this abrupt increase the
$S$-parameter hardly changes, although the bar strength increases considerably.
 
%--------------------------------------------------------------------
%--------------------------------------------------------------------

\subsection{Secular evolution of the halo component}
\label{subsec:halobar}

The halo also undergoes some secular evolution, albeit not as strong as that of
the disk. The most notable feature is that an initially axisymmetric halo
becomes elongated in its innermost parts and forms what is usually called the
`halo bar', or the `dark matter bar', although the word `bar' in this context is
rather exaggerated, and `oval' would have been more appropriate. This structure
was already observed in a number of simulations (e.g., Debattista \& Sellwood
2000; O'Neil \& Dubinski 2003; Holley-Bockelmann \textit{et al.} 2005; Berentzen
\& Shlosman 2006) and its properties have been studied in detail by Hernquist \&
Weinberg (1992), Athanassoula (2005a, 2007) and Colin \textit{et al.} (2006). It
is considerably shorter and its ellipticity is much smaller than the disk bar,
while rotating with roughly the same angular velocity. It is due to the
particles in the halo ILR (Athanassoula 2003, 2007). 

A less clear-cut and certainly much more debated issue concerns the question
whether secular evolution due to a strong bar could erase the cusp predicted by
cosmological simulations and turn them into cores, which would lead to an
agreement with observations. A few authors (e.g., Hernquist \& Weinberg 1992;
Weinberg \& Katz 2002; Holley-Bockelmann \textit{et al.} 2005; Weinberg \& Katz
2007a, 2007b) argued that indeed such a flattening was possible, while a larger
consensus was reached for the opposite conclusion (e.g., Sellwood 2003; McMillan
\& Dehnen 2005; Colin \textit{et al.} 2006; Sellwood 2008). We refer the reader to these
papers for more information.   

%
%%%%%%%%%%%%%%%%%%%%%%%%%%%%%%%%%%%%%%%%%%%%%%%%%%%%%%%%%%%%%%%%%%%%%%%%%%%%%%%%
%

\section{Comparison with observations}
\label{sec:observations}

Technically, comparison between observations and simulations is relatively
straightforward. From the coordinates of the particles in the luminous
components, and after choosing the viewing angles and taking into account the
observational conditions, it is possible to obtain an image that can be output
in the standard format used by observers, namely FITS (Flexible Image Transport
System). This image can then be analysed as are observations, using standard
packages, such as IRAF (Image Reduction and Analysis Facility). Similarly one
can create data cubes, containing velocity information, which again will be
analysed with the same software packages as observations. Taking into account
the limitations of the instruments and more generally those due to observational
conditions is an important feature here, as is the fact that it is the
simulation data that must be transformed into observations and not the other way
round.

There is, nevertheless, one subtle point concerning a limitation of dynamical
simulations that should  be kept in mind. It concerns the simulation time to be
chosen for the comparison. As already mentioned, in dynamic simulations the disk
is assumed to be in place and in equilibrium before the bar starts forming. On
the contrary, in cosmological simulations the bar should start forming as soon
as the relative disk mass is sufficiently high to allow the bar instability to
proceed. One must add to this the uncertainty about when disks can be considered
as being in place. All this taken together makes it very difficult to pinpoint
the simulation time to be used for the comparisons. The best is to try a range
of times and then describe how the fit evolves with time.

%
%%%%%%%%%%%%%%%%%%%%%%%%%%%%%%%%%%%%%%%%%%%%%%%%%%%%%%%%%%%%%%%%%%%%%%%%%%%%%%%%
%

\section{Summary and discussion}
\label{sec:summary}

Angular momentum can be redistributed within a barred galaxy. It is emitted
from the (near-)resonant stars in the bar region and absorbed by the
(near-)resonant material in the spheroid and the outer disk. By following the
orbits in a simulation and measuring their frequencies, it is possible to
determine whether they are (near-)resonant or not, and, if so, at which
resonance. For strong bar cases, the most populated disk resonance is the inner
Lindblad resonance. Simulations confirm the theoretical prediction that this
emits angular momentum, and that the corotation and outer Lindblad resonances
absorb it. In the spheroid the three most populated resonances are the
corotation, the outer Lindblad and the inner Lindblad resonance, and, in many
cases, it is corotation that is the most populated. Again simulations confirm
the theoretical prediction that angular momentum is absorbed at the spheroid
resonances. 

In order for bars to evolve uninhibited in a simulation, it is necessary that
the angular momentum exchange is not artificially restrained, as would be the
case if the halo in the simulation was rigid, e.g., represented by an
axisymmetric force incapable of emitting or absorbing angular momentum. It is
thus necessary to work with live haloes in simulations, and, more generally, to
avoid the use of any rigid component.    

Note also that the effect of the spheroid on bar growth is different in the
early and in the late phases of the evolution. During the initial phases of the
evolution, the spheroid, due to the strong axisymmetric force it exerts, delays
and slows down the bar growth. Thus, {\it bars will take longer to form in
galaxies with a large ratio of spheroid-to-disk mass}. On the other hand, at
later stages, after the secular evolution has started, the spheroid can increase
the bar strength by absorbing a large fraction of the angular momentum emitted
from the bar region. Thus, {\it stronger bars will be found in galaxies with a
larger spheroid-to-disk mass ratio}. 

Contrary to spheroid mass, the velocity dispersion in the disk has always the
same effect on the bar growth. During the initial phases it slows down the bar
growth. Thus, {\it bars will take longer to form in galaxies with hot disks}.
During the secular evolution phase, a higher velocity dispersion in the disk
component will make its resonances less active, since it decreases the amount of
angular momentum that a resonance can emit or absorb. A similar comment can be
made about the velocity dispersion of the spheroid (near-)resonant material.
Thus, {\it increasing the velocity dispersion in the disk and/or the spheroid 
will lead to less angular momentum redistribution and therefore weaker bars}. 
 
As the bar loses angular momentum, its pattern speed decreases, so that the
resonant radii will   move outwards with time. Since the corotation radius
provides an absolute limit to the bar length, this increase implies that the bar
can become longer. Indeed, this occurs in simulations. It is thus possible for
the pattern speed to decrease while the bar stays `fast', provided the bar
becomes longer in such a way that the ratio $\mathcal{R}$ of corotation radius
to bar length stays within the bracket 1.2\,$\pm$\,0.2.

As the bar loses angular momentum it also becomes stronger, so that there is a 
correlation between the bar strength and the amount of angular momentum
absorbed by the spheroid. In general, as bars become stronger they become also
longer and their shape gets more rectangular-like. They redistribute mass within
the disk and create the disky bulge (more often referred to as pseudo-bulge) in
the central region. They also increase the disk scalelength. All these changes
brought about by the evolution can also strongly influence the form of the
rotation curve and change an initially sub-maximum disk to a maximum one. 

The strongest bars will be found in cases where the maximum amount of angular
momentum has been redistributed within the galaxy, and not when the spheroid
mass is maximum. A further parameter which is crucial in trying to maximise the
angular momentum redistribution is the bar pattern speed. Indeed, this is set by
the location of the corotation radius and therefore by the balance between
emitters and absorbers in the disk. 

When bars form they are vertically thin, but soon their inner parts puff up and
form what is commonly known as the boxy/peanut bulge. This is well understood
with the help of orbital structure theory. It gives a complex and interesting
shape to the bar -- i.e., vertically extended only over a radial extent from the
centre to a maximum radius of the order of (0.7\,$\pm$\,0.3)$a_{\rm B}$, where
$a_{\rm B}$ is the bar length, and then very thin outside that range. This shape
explains a number of observations and also argues that the \textit{COBE}/DIRBE
bar and the Long bar in our Galaxy are, respectively, the thin and the thick
part of a single bar.

From the above it is thus possible to conclude that there is a continuous
redistribution of angular momentum in disks with strong bars and that this 
drives a secular evolution. It is secular because the timescales involved are
long, contrary to, e.g., a merging, which occurs in a very short time interval. 

%
%%%%%%%%%%%%%%%%%%%%%%%%%%%%%%%%%%%%%%%%%%%%%%%%%%%%%%%%%%%%%%%%%%%%%%%%%%%%%%%%
%
\section*{Acknowledgments}
I thank the school organisers, J. Falc\'on-Barroso and J.~H.~Knapen for inviting
me to give a series of lectures at the XXIIIrd Canary Islands Winter School on
Secular Evolution of Galaxies, as well as for their patient nudging when the
time came to write up the proceedings. I acknowledge financial support from the
CNES and from the People Programme  (Marie Curie Actions) of the European
Union's FP7/2007-2013/ to the DAGAL network under REA grant agreement number
PITN-GA-2011-289313. 

%
%%%%%%%%%%%%%%%%%%%%%%%%%%%%%%%%%%%%%%%%%%%%%%%%%%%%%%%%%%%%%%%%%%%%%%%%%%%%%%%%
%

\end{document}